\documentclass[11pt]{article}
\usepackage{graphicx} 

\usepackage{natbib}
\usepackage{times}
\usepackage{amsmath}
\usepackage{amssymb}
\usepackage{amsthm}
\usepackage{xcolor}
\usepackage{thm-restate}
\usepackage{mathtools}
\usepackage{thmtools}
\usepackage{enumitem}
\usepackage{circledsteps}
\usepackage{subcaption}
\usepackage{booktabs}
\usepackage{ifthen}
\newboolean{isSingleColumn}

\newcommand{\cS}{\mathcal{S}}
\newcommand{\cM}{\mathcal{M}}
\newcommand{\vx}{\mathbf{x}}
\newcommand{\vy}{\mathbf{y}}
\newcommand{\vf}{\mathbf{f}}

\newcommand{\vu}{\mathbf{u}}
\newcommand{\vb}{\mathbf{b}}
\newcommand{\vpi}{\mathbf{\pi}}
\newcommand{\veta}{\mathbf{\eta}}
\newcommand{\vlambda}{\mathbf{\lambda}}
\newcommand{\vgamma}{\mathbf{\gamma}}
\newcommand{\milp}{\texttt{MILP}}
\newcommand{\infmilp}{\texttt{infMILP}}
\newcommand{\csuccW}{s \in \cS \mid c \succ_s W}
\newcommand{\supp}{\operatorname{supp}}

\newtheorem{corollary}{Corollary}
\newtheorem{theorem}{Theorem}
\newtheorem{example}{Example}
\newtheorem{definition}{Definition}

\usepackage{empheq}
\usepackage{tikz}
\usepackage{bbding}
\usetikzlibrary{arrows.meta}

\usepackage{nicefrac}
\usepackage{tabularx}
\usepackage{bbm}

\newcolumntype{Y}{>{\centering\arraybackslash}X}

\makeatletter
\if@twocolumn
    \setboolean{isSingleColumn}{false}
\else
    \setboolean{isSingleColumn}{true}
\fi

\newcommand{\adaptiveHeader}[1]{%
    \ifthenelse{\boolean{isSingleColumn}}{%
        \paragraph{#1} 
    }{%
        \textbf{#1}    
    }%
}

\newcommand{\tabwidth}{\linewidth}

\usepackage{fullpage}
\usepackage{authblk}

\usepackage{hyperref}
\usepackage{cleveref}

\title{MILP the Gap: An Automated Reasoning\\ Approach to Condorcet's Paradox}

\author[1]{Itai Zilberstein\thanks{Correspondence to \texttt{izilbers@cs.cmu.edu}.}}
\author[1]{Ruben Martins}
\author[1,2]{Ratip Emin Berker}

\affil[1]{Department of Computer Science, Carnegie Mellon University, Pittsburgh, PA}
\affil[2]{Foundations of Cooperative AI Lab (FOCAL)}

\begin{document}

\maketitle
\thispagestyle{empty}

\begin{abstract}
Computational social choice studies how conflicting individual preferences are aggregated into a collective decision, with applications to elections, AI evaluation/alignment, and recommendation systems. We focus on settings where $n$ \emph{voters} (\textit{e.g.}, human evaluators or algorithmic benchmarks) each rank $m$ \emph{alternatives} (\textit{e.g.}, human candidates or AI models/outputs), and the goal is to select a \emph{committee} of $k$ alternatives. One measure of the (in)\emph{stability} of the outcome is the number of voters who prefer an unchosen alternative to all alternatives in the committee.  If this number is less than half the electorate, the committee is a \emph{Condorcet winning set}, which may not exist if $k \leq 2$. More generally, there exist instances where any committee of size $k$ has at least a $\nicefrac{2}{(k+1)}-\epsilon$ fraction of the electorate preferring another alternative, for arbitrarily small $\epsilon$. While recent work has made remarkable progress narrowing the gap, it is currently unknown whether this lower bound is tight.

We develop a \emph{mixed-integer linear programming} framework that computes exact bounds for fixed parameter values. Our approach depends only on $m$ and $k$; by modeling electorates as distributions over rankings, we eliminate any dependence on $n$. This provides a middle ground between theoretical approaches that seek to give bounds depending solely on $k$, which are general but currently not tight, and previous SAT-based approaches, which scale poorly with $n$ and cannot establish generalizable bounds.  Our programs allow us to rule out counterexamples over arbitrary electorate sizes and obtain tight bounds for fixed $m$ and $k$. Further, our framework provides a pathway for proving general upper bounds. In particular, we show a linear regime, $k \ge \nicefrac{8m}{9}$, in which the conjectured $\nicefrac{2}{(k+1)}$ upper bound indeed holds, providing the first results that match known lower bounds. Beyond stability, our framework can characterize the relationships among other social choice axioms, which we demonstrate by producing minimal incomparability results. 
\end{abstract}

\clearpage
\setcounter{page}{1}

\section{Introduction}\label{sec:intro}

Computational social choice provides a formal framework for studying how individual preferences can be aggregated into collective decisions. Preference aggregation underlies many high-stakes decision-making processes in today's world, including elections for representative governments~\citep{Arrow51:Social}, AI evaluation and alignment~\citep{Conitzer24:Position}, recommendation systems~\citep{Pennock00:Social,Skowron16:Finding}, and scarce resource allocation~\citep{Freedman20:Adapting,Zilberstein26:Preference}. Across these settings, a central challenge is to identify collective outcomes that accommodate conflicting preferences.

In this paper, we consider elections in which $n$ \emph{voters} each rank $m$ \emph{alternatives}. These terms encompass more than political elections: voters may represent human evaluators or benchmark tasks, while alternatives may represent candidates for office or employment, recommended items, or AI models and their outputs. 

A classic obstacle in ranked choice voting is the \emph{Condorcet paradox}~\citep{Condorcet14:Essai}: there are elections in which selecting any single alternative as the winner results in a majority of voters preferring an alternative. We illustrate such an election in~\Cref{tab:condorcet-paradox}. In the worst case over all elections, the fraction of the electorate that prefers another alternative over the selected winner can come arbitrarily close to one, rendering single-winner systems theoretically unstable. 

Selecting a \textit{committee} of $k$ alternatives is a natural relaxation of the single-winner setting. Committee selection also arises in many real-world domains (\textit{e.g.}, shortlisting multiple job applicants for interviewing or electing parliamentary representatives). Here, a committee's \emph{stability} can be measured analogously to a single winner through the largest fraction of voters who prefer any single unselected alternative to every committee member. A committee is called a \emph{Condorcet winning set (CWS)} if this fraction is less than one half~\citep{Elkind11:Choosing,Elkind15:Condorcet}. This raises a natural question: how large must a committee be to guarantee this form of stability, regardless of the number of alternatives, the number of voters, or their preferences?

This question was first posed in 2015 by ~\citet{Elkind15:Condorcet}, who established that a CWS of size $O(\log m)$ always exists. On the negative side, they gave an election where no two-member committee is a CWS. More generally, their construction can be extended to show that for every $\alpha< \nicefrac{2}{(k+1)}$, there are elections in which every committee of size $k$ faces an unselected alternative that at least an $\alpha$ fraction of voters prefer to every committee member \cite{Charikar25:Six}. The gap remained open for almost a decade, until \citet{Charikar25:Six} showed six alternatives always suffice for a CWS, independently of the number of alternatives or voters, and \citet{Song26:Few} improved this bound to five. More generally, ideas from the latter can be used in tandem with prior proof techniques \cite{Jiang20:Approximately, Charikar25:Six} to prove the existence of a $k$-member committee where no other alternative is preferred by a $\approx \nicefrac{4.910}{k}$ fraction of the voters \cite{Charikar26:Exposition}. Despite these advances, it remains one of the most elusive questions in social choice as to whether three or four alternatives always suffice to form a CWS, and, more generally, whether the lower bound of $\nicefrac{2}{(k+1)}$ is in fact tight, as conjectured by \citet{Charikar25:Six}. Closing this gap requires either stronger existence guarantees or elections that demonstrate their impossibility. 

\begin{table}[t]
    \centering
    \caption{An election with six candidates where for any chosen winner, another alternative is preferred by five out of the six voters~\cite{Charikar25:Six}. Each column shows one voter ranking, with the top candidate ranked first, \textit{e.g.}, voter 1 ranks $c_1 \succ c_2 \succ \ldots \succ c_6$.}
    \label{tab:condorcet-paradox}
    \begin{tabular}{c c c c c c}
         $\succ_1$ & $\succ_2$ & $\succ_3$ & $\succ_4$ & $\succ_5$ & $\succ_6$ \\
         \hline
         $c_1$ & $c_2$ & $c_3$ & $c_4$ & $c_5$ & $c_6$ \\
         $c_2$ & $c_3$ & $c_4$ & $c_5$ & $c_6$ & $c_1$ \\
         $c_3$ & $c_4$ & $c_5$ & $c_6$ & $c_1$ & $c_2$ \\
         $c_4$ & $c_5$ & $c_6$ & $c_1$ & $c_2$ & $c_3$ \\
         $c_5$ & $c_6$ & $c_1$ & $c_2$ & $c_3$ & $c_4$ \\
         $c_6$ & $c_1$ & $c_2$ & $c_3$ & $c_4$ & $c_5$
    \end{tabular}
\end{table}

\paragraph{Our contributions}

We reframe worst-case committee stability as an optimization problem over election instances
and develop a framework based on \textit{mixed-integer linear programming (MILP)} that computes exact worst-case values of $\alpha$ for specific parameters. By modeling electorates as distributions over rankings, our formulation depends only on the number of alternatives $m$ and committee size $k$, not on the number of voters $n$. We develop two main programs with different advantages. The first program, \milp, computes the exact worst-case value of $\alpha$ for fixed $m$ and $k$. We develop a number of optimizations to scale \milp, such as symmetry breaking, and we prove the correctness of these methods.  We computationally establish that, in every election with less than nine alternatives and a committee size greater than two, a CWS exists. More generally, we establish that every election with less than nine alternatives satisfies the conjectured upper bound of $\nicefrac{2}{(k+1)}$. 

Our second program, referred to as \infmilp, uses an enhanced formulation; we treat each alternative as a type representative of an infinite cycle of \emph{clones}. We prove theoretical lower bounds for both \milp~(\Cref{prop:milp-lower-bound}) and \infmilp~(\Cref{prop:infmilp-lower-bound}). In contrast to \milp, \infmilp~recovers the known general lower bound for any parameter assignment, providing incumbent solutions at the edge of established values. 

Complementing these computational methods, we provide an explicit upper bound on \milp. We prove that every election has a committee in which every unelected alternative is supported by less than a $\allowbreak \min_{0 \leq t \leq k}\frac{m-t}{(t+1)(k-t+1)}$ fraction of the electorate (\Cref{thm:milp-upper-bound}). Our upper bound can be used to show that $\alpha \leq \nicefrac{2}{(k+1)}$ whenever $k \ge \nicefrac{8m}{9}$ (\Cref{cor:linear_regime}), providing the first result that establishes the conjectured threshold in a linear parameter regime. 

Finally, we extend our automated reasoning framework to characterize the relationship between committee stability and other proportionality axioms. We obtain the minimum-sized counterexamples establishing that stability is incomparable with three axioms: rank-priceability, Rank-PJR+, and Rank-EJR+~\citep{Brill23:Robust}. We also use our framework to computationally verify that in every election with at most six alternatives, one proportionality axiom, namely Rank-PJR+, and stability are always simultaneously achievable. 

By introducing a program that is able to compute tight values of $\alpha$ for fixed $m$ and $k$ (and over all possible $n$), our approach provides a middle ground between theoretical approaches that seek to give bounds depending solely on $k$, which are general but currently not tight (\textit{e.g.},~\citep{Charikar25:Six,Song26:Few}), and previous SAT-based approaches, which scale poorly with $n$ and cannot establish generalizable bounds (\textit{e.g.},~\citep{Geist16:Findingp}). By automating the search for counterexamples (and thus ruling out their existence in specific regimes) our framework also complements the increasingly successful use of generative AI to assist in tackling open conjectures in computational social choice (\textit{e.g.},~\citep{Manurangsi26:Polynomial,Becker26:Existence,Frank26:Polynomial}). Unlike searching for counterexamples (or certificates of their nonexistence) via generative AI, which may at times lead to hard-to-verify proofs~\citep{Boer26:OpenAI,Chang26:OpenAI}, our automated reasoning approach exhaustively searches over the space of constructions and produces independently checkable certificates. These outputs are easily verifiable and can be immediately translated into concise, human-readable proofs. Our minimum-size counterexamples of incomparability from \Cref{sec:axioms} illustrate this.

\section{Related work}\label{sec:related}

\paragraph{Committee selection and proportionality} Selecting a committee of candidates that proportionally represents the preferences of an electorate is a classic problem in social choice~\citep{Droop81:Methods}. When voters provide preferences via rankings of candidates, the concept of a Condorcet winning set (CWS) as a stable set of winners was formalized by~\citet{Elkind11:Choosing}. This led to the notion of the \textit{Condorcet dimension} of an election: the smallest $k$ such that a size $k$ CWS exists~\citep{Elkind15:Condorcet}. The Condorcet dimension has been extended to other settings beyond ranked choice voting, such as elections in which voter preferences are measured by distances to candidates~\citep{Lassota25:Condorcet}. 

The generalization of CWS to an arbitrary fraction of voters is closely linked to the literature on \emph{stable committees}~\cite{Jiang20:Approximately,Cheng19:Group}, where a committee is said to be \emph{core-stable} (or \emph{in the core}) if no $x\%$ of the electorate can all be better off with another subcommittee that only uses $x\%$ of the seats. In the ranking setting (where each voter's utility is determined by their favorite candidate in the committee), this is equivalent to requiring that there be no alternative that is preferred by at least $\nicefrac{1}{k}$ fraction of the voters (or strictly greater than $\nicefrac{1}{(k+1)}$, using the more stringent \emph{Droop quota}~\cite{Brill20:Approval}). While the existing counterexamples showing the $\nicefrac{2}{(k+1)}$ lower bound rule out the existence of core-stable committees in this setting, upper bounds prove the existence of \emph{approximately} stable committees.

Beyond committee stability, prior literature has also proposed other axioms capturing proportionality and fair representation in multi-winner elections~\citep{Aziz17:Condorcet}. \citet{Brill23:Robust} introduced Rank-PJR+, Rank-EJR+, and rank-priceability, extending these proportionality notions to ranked preferences from the approval-based setting; the latter builds on approval-based priceability~\citep{Peters20:Proportionality}. 



\paragraph{Automated reasoning and social choice} There is a growing body of work leveraging automated reasoning to prove or disprove conjectures in social choice theory~\cite{Geist17:Computer}. Early work successfully applied automated reasoning to rederive impossibility theorems~\cite{Tang08:Computer,Tang09:Computer}, such as that of Arrow~\cite{Arrow51:Social} and Gibbard–Satterthwaite~\cite{Gibbard73:Manipulation,Satterthwaite75:Strategy}. Further encoding of social choice problems into SAT led to results in a variety of settings~\cite{Geist11:Automated,Brandt16:Finding,Brandt17:Optimal,Brandl19:Strategic,Brandl21:Distribution}. Most relevant to our work is \citet{Geist16:Findingp}, who encoded the search for CWS as a satisfiability problem. While they successfully identified minimal elections with Condorcet dimension three, their approach faced significant computational hurdles. The SAT encoding requires fixing the number of candidates \emph{and voters} in advance, leading to poor scalability (the biggest elections tested by \citet{Geist16:Findingp} contain $n=12$ voters).

More recently, \emph{Satisfiability Modulo Theories (SMT)} and \textit{mixed integer linear programming (MILP)} formulations have seen success in social choice, particularly in probabilistic settings~\cite{Mennle16:Pareto,Brandl18:Proving}. Of particular relevance is the work of \citet{Berker25:Edge}, who successfully applied MILP techniques to prove bounds on core stability in approval-based multi-winner voting.\footnote{Since then, the existence of core-stable committees in that setting has been proven~\citep{Becker26:Existence}.} Our automated reasoning approach in the ranked setting complements their work in the approval-based setting. As they did, we relax the integrality of the voters, modeling the electorate as a probability distribution over rankings. This allows us to search the space of elections with an effectively \emph{infinite} number of voters, removing the prior dependency of SAT-based approaches. We also take a similar approach to investigate the relationship among different axioms.

Another successful application of MILP is in the field of automated mechanism design~\citep{Conitzer02:Complexity,Conitzer04:Self}. In this setting, automated reasoning proved both impossibility and existence results, such as for redistribution mechanisms~\citep{Guo09:Worst,Guo10:Computationally,Damle24:Designing}. 
\section{Preliminaries}

A ranking-based election consists of $m$ alternatives (or candidates) $C = \{c_1, \dots, c_m\}$ and $n$ voters $N = \{1, \dots, n\}$. Each voter $i \in N$ has a strict ranking $\succ_i$ over $C$. In multi-winner ranked voting, the goal is to select a committee $W \subset C$ of size $k$, where $0 < k < m$.  We denote the set of all such committees by $\cM_k = \{W \subseteq C \mid |W| = k\}$.

\begin{definition}[$\alpha$-undominated set]
A set of alternatives $W \subseteq C$ is an $\alpha$-undominated set if, for every alternative $c \in C \setminus W$, strictly less than an $\alpha$ fraction of the voters prefer $c$ over every member of $W$.
\end{definition}

A \textit{Condorcet winning set (CWS)} is an $(\nicefrac{1}{2})$-undominated set.  

To study the worst-case stability over any number of voters, we represent the electorate as a distribution over rankings rather than enumerating individual voters as above. Let $\cS$ be the set of all $m!$ rankings of the alternatives $C$. Every election induces a probability distribution $\vx \in \Delta(\cS)$, where $\vx[s]$ is the fraction of voters with the ranking $s \in \cS$. Similarly, any voter distribution with rational inputs is induced by some election with sufficiently many (integral) voters. For a ranking $s$, we write $c \succ_s c'$ if alternative $c$ is ranked higher than $c'$ according to $s$ and $c \succ_s W$ if alternative $c$ is ranked higher than each $c' \in W$ according to $s$. One useful aspect of this formulation is that the fraction of voters who prefer an alternative $c$ over a committee $W$ according to profile $\vx$ is given by $\sum_{s \in \cS \mid c \succ_s W} \vx[s]$. Then, $W$ is an $\alpha$-undominated set if and only if
\begin{equation}
\label{eq:alpha-inequality}
\forall c \in C \setminus W  \quad \smashoperator{\sum_{\csuccW}} \vx[s] < \alpha.
\end{equation}

We use this characterization to formulate worst-case committee stability as an optimization problem parameterized by $m$ and $k$.



\section{Stability as an optimization problem}
\label{sec:milps}

In this section, we develop mixed-integer linear programs (and computational speedups to them) that compute voter profiles for which the \emph{most stable} committee is \emph{as unstable as possible}. In other words, for fixed $m$ and $k$, we compute the worst-case $\alpha^*$, which characterizes the threshold above which an $\alpha$-undominated set of size $k$ is guaranteed to exist for every voter profile over $m$ candidates.

To guarantee the existence of an $\alpha$-undominated set of size $k$ in any election with $m$ alternatives, there must exist a committee $W$ for every voter profile such that Inequality \ref{eq:alpha-inequality} holds. In other words,
\begin{equation*}
\forall \vx \in \Delta(\cS), ~\exists W \in \cM_k, ~\forall c \in  C \setminus W \quad \smashoperator{\sum_{\csuccW}}  \vx[s] < \alpha.
\end{equation*}

We can then formulate the worst-case margin $\alpha^*$ for committees of size $k$ by the following optimization problem. 

\begin{equation}\label{eq:alphastar}
\alpha^* = \max_{\vx \in \Delta(\cS)}  \min_{W \in \cM_k}  \max_{c \notin W} \quad \smashoperator{\sum_{\csuccW}} \vx[s]. \tag{$\alpha\texttt{-Opt}$}
\end{equation}

This optimization problem exactly characterizes the threshold above which an $\alpha$-undominated committee of size $k$ is guaranteed to exist in every election with $m$ alternatives.

\begin{restatable}{proposition}{OptCorrectness}
\label{prop:opt-correctness}
Suppose $m$ and $k$ are fixed and let $\alpha^*$ be the optimal value of~\ref{eq:alphastar}. Then an $\alpha$-undominated set of size $k$ is guaranteed to exist for every voter profile if and only if $\alpha > \alpha^*$.
\end{restatable}

The proof of~\Cref{prop:opt-correctness}, along with all other omitted proofs in this section, is provided in~\Cref{sec:ommitted-proofs}. 

Equivalently to~\Cref{prop:opt-correctness}, there exists a finite set of voters for which no $\alpha$-undominated set of size $k$ exists if and only if $\alpha \leq \alpha^*$. We can interpret the optimization problem as a two-player zero-sum game. The \textit{maximizer} selects a voter profile $\vx$, as well as a mapping from each committee $W$ to its strongest challenger maximizing the sum in \eqref{eq:alphastar}; this mapping effectively determines the inner maximization for each value of $W$. The \textit{minimizer}, in response, selects a committee $W$ with the goal of minimizing this sum. 

To guarantee the existence of a CWS for a given $m$ and $k$, one would need to assert that $\alpha^* < \nicefrac{1}{2}$.  On the other hand, to disprove the existence of a CWS of size $k$, one just needs to show that $\alpha^* \ge \nicefrac{1}{2}$ for some $m$ and $k$. 

\begin{corollary}
    Suppose $m$ and $k$ are fixed and let $\alpha^*$ be the optimal value of~\ref{eq:alphastar}. If $\alpha^* \ge 1/2$, then there exist $n$ voters for which no CWS of size $k$ exists.
\end{corollary}

\subsection{A mixed-integer program for stability}

We can further reformulate the optimization problem in~\eqref{eq:alphastar} as a mixed-integer linear program, referred to as \milp. The program takes as input the alternative count $m$ and the committee size $k$.

\begin{align*}
    \max \quad & \alpha  \\
    \text{s.t.} \quad & \sum_{s\in\cS} \vx[s] = 1 \\
    & \sum_{c\notin W} \vy[W,c] = 1, \quad \forall W\in \cM_k \\
    & \smashoperator{\sum_{\csuccW}} \vx[s] + Q(1-\vy[W,c]) \ge \alpha, \quad \forall W\in \cM_k, c\notin W \\
    & \vx \ge 0, \quad \vy[W,c]\in \{0, 1\}, \quad \alpha \ge 0 
\end{align*}

 Here, $Q$ is a constant we set to be $Q \geq 1$. The variable $\vx$ is a vector indexed by the set of voter rankings $s \in \cS$. The constraints enforce that $\vx$ is a valid probability distribution. The variable $\vy[W,c]$ is a binary variable that indicates whether $c$ is the alternative chosen to challenge $W$. Each committee $W\in \cM_k$ must have exactly one alternative challenging it. Finally,  we enforce the max-min logic by bounding  $\alpha$ from above by the fraction of voters supporting challenger $c$ against $W$ when $\vy[W,c] = 1$. By setting $Q\ge1$, we effectively relax the bound on $\alpha$ for committees and alternatives where $\vy[W,c] = 0$. 

\begin{restatable}{theorem}{MILPCorrectness}
\label{thm:milp-correctness}
    Fix any $m$ and $k$. Let $\alpha_1^*$ be the optimal value of~\ref{eq:alphastar} and $\alpha_2^*$ be the optimal value of \milp. Then, $\alpha_1^* = \alpha_2^*$. Moreover, a profile $\vx^*$ is optimal for~\ref{eq:alphastar} if and only if it can be extended with variables $\vy$ to an optimal solution of \milp.
\end{restatable}

At an optimal solution, $\vy[W,c]=1$ identifies a challenger whose support gives the lower bound $\alpha^*$. The final $\vx^*$ also gives the voter profile that achieves the largest value of $\alpha$. Since all the coefficients of \milp~are rational, there exists an optimal solution $\vx^*$ that is also rational. 

The complexity of \milp~ is manageable at small values of $m$ and $k$, but scales factorially with $m$ due to the dependency on $\cS$, which has size $\mathcal{O}(m!)$. The vector $\vy$ also adds an additional $(m-k)\cdot {m \choose k}$ binary variables. For this reason, we use several modeling tricks to improve the efficiency of the program. 

We can reduce the size of $\cS$ by observing that the specific relative ordering of alternatives in the bottom $k$ positions of a ranking does not affect any of the constraints. Consider two cases.
\begin{enumerate} 
\item If a challenger $c$ is in the bottom $k$ positions of a ranking, and the committee $W$ has size $k$, then by the Pigeonhole Principle, at least one member of $W$ must reside in the top $m-k$ positions. Thus, $W$ defeats $c$ in this ranking regardless of the specific permutation of the bottom $k$ alternatives.
\item If a challenger $c$ is in the top $m-k$ positions, it defeats any committee member located in the bottom $k$ positions, regardless of how these $k$ positions are ordered. To determine if $c \succ_s W$, we only need to compare $c$ against the committee members located in the top $m-k$ positions.
\end{enumerate} 

Consequently, we can group all $m!$ rankings into equivalence classes according to their ordered top-$(m-k)$ prefix. There are $m!/k!$ such prefixes. With a slight abuse of notation, we let $\cS$ denote this reduced set of ordered prefixes and let $\vx[s]$ denote the probability mass of all complete rankings in the corresponding equivalence class for \milp~ in this subsection only. 

The election structure also contains many isomorphisms. For instance, permuting alternative labels does not change the solution. We can exploit this symmetry by fixing the committee and challenger that attain the minimum and maximum, respectively, in~\ref{eq:alphastar}. Consider an optimal profile $\vx^*$, let

$$ W^* \in \arg\min_{W\in\cM_k} \max_{c\notin W} \smashoperator{\sum_{\csuccW}}\vx^*[s], \text{ and} $$

$$c^* \in \arg\max_{c\notin W^*} \smashoperator{\sum_{s\in\cS \mid c\succ_s W^*}}\vx^*[s]. $$

Because alternative labels are arbitrary, we may assume without loss of generality that $W^*=\{c_1,\ldots,c_k\}$ and $c^*=c_{k+1}$. At an optimal solution, $\alpha^* = \sum_{s\in\cS \mid c_{k+1}\succ_s W^*}\vx^*[s]$. We obtain the following formulation of \milp~ with this symmetry broken.

\begin{align*}
    \max \quad & \alpha  \\
    \text{s.t.} \quad & \sum_{s\in\cS} \vx[s] = 1 \\
    & \sum_{c\notin W} \vy[W,c] = 1, \quad \forall W\in \cM_k \setminus \{W^*\}\\
    & \smashoperator{\sum_{\csuccW}} \vx[s] + (1-\vy[W,c]) \ge \alpha,  \forall W\in \cM_k \setminus \{W^*\}, c\notin W \\
    & \smashoperator{\sum_{s \in \cS \mid c_{k+1} \succ_s W^*}} \vx[s]  = \alpha \\
    & \vx \ge 0, \quad \vy[W,c]\in \{0, 1\}, \quad \alpha \ge 0 
\end{align*}

Compared with the original program, we no longer require variables for $\vy[W^*,c]$ for any $c\in C$. Instead, we fix $c_{k+1}$ as the challenger against $W^*$ and force $c_{k+1}$ to have support exactly $\alpha$. This version of \milp~ with symmetry broken preserves the correctness.

\begin{restatable}{proposition}{MILPSymmetryCorrectness}
\label{prop:symmetry-milp}
For fixed $m$ and $k$, the symmetry-broken formulation of \milp~ has the same optimal value as~\ref{eq:alphastar}.
\end{restatable}

\subsection{A clone-enhanced mixed-integer program}

In this section, we enhance \milp~ using a conceptual trick that will ultimately let us produce stronger lower bound constructions via a compact representation of structured elections with arbitrarily many alternatives.  We extend our search to a family of elections in which each of the $m$ alternatives in the original representation is now interpreted as an alternative \emph{type}. Each alternative type represents a block of cyclically ranked \emph{clones},\footnote{In ranking-based elections, clones refer to a set of candidates ranked as an interval by all voters~\citep{Tideman87:Independence}. Clone structures, as well as their impact on results under various voting rules, have been well-studied in the social choice literature (\textit{e.g.}, \citep{Elkind10:Clone,Brandl16:Consistent,Berker25:From}).} with each rotation appearing equally often among voters. Formally, fix an integer $K>k$, and replace each alternative type $c_i\in C$ by $K$ distinct alternatives $c_i^{(1)},\ldots,c_i^{(K)}$. For each ranking $s$, the original alternative types are all ordered according to $s$. Within each type, the $K$ cyclic rotations
\[ c_i^{(j)} \succ c_i^{(j+1)} \succ \cdots \succ c^{(K)}_i \succ c_i^{(1)} \succ \cdots \succ c_i^{(j-1)}, \quad j\in[K], \]
occur with equal probability, independent of the ranking. The cyclic structure forces the optimization to find committees that are robust even to challengers drawn from the same alternative type as a committee member. For example, the election from \Cref{tab:condorcet-paradox} can be thought of as an election with a single type consisting of $K=6$ clones. Indeed, this structured family of elections was used by \citet[Appendix B]{Charikar25:Six} to prove the $\nicefrac{2}{(k+1)}$ lower bound.

Taking the number of clones per type to infinity yields the new program, \infmilp. This allows us to search for lower bound constructions that involve arbitrarily many alternatives while optimizing the ranking of only $m$ types. Like \milp, the program \infmilp~takes as input the number of alternative types $m$ and the committee size $k$. However, we replace $\cM_k$ by the set of all $k$-multisets over $C$, denoted $\cM_k^+$. This is because multiple copies of the same alternative type may now appear in a committee, representing the selection of multiple distinct clones of the same type. Likewise, a committee may be challenged by an unselected clone of a type already represented in the committee. To build intuition for this construction, we provide a simple example with three clones.

\begin{example} We illustrate an election with a cycle of clones in~\Cref{fig:infmilp-cycle-example}.  There is a single alternative $\textcolor{red}{c}$ with three clones, $\textcolor{red}{c^{(1)}}, \textcolor{red}{c^{(2)}}$, and $\textcolor{red}{c^{(3)}}$. The electorate is divided equally among the three cyclic rankings: $\textcolor{red}{c^{(1)}} \succ \textcolor{red}{c^{(2)}}\succ \textcolor{red}{c^{(3)}}$, $\textcolor{red}{c^{(2)}} \succ \textcolor{red}{c^{(3)}}\succ \textcolor{red}{c^{(1)}}$, and $\textcolor{red}{c^{(3)}} \succ \textcolor{red}{c^{(1)}}\succ \textcolor{red}{c^{(2)}}$.  No single clone can form a CWS. The committee must contain multiple distinct copies of $\textcolor{red}{c}$ to be stable. The multiset committee $W = \{\textcolor{red}{c}, \textcolor{red}{c}\}$ (instantiated concretely as $\{\textcolor{red}{c^{(1)}}, \textcolor{red}{c^{(2)}}\}$) is a CWS. 
\end{example}

\begin{figure}
\centering
\scalebox{0.9}{
\begin{tikzpicture}[
    main node/.style={
        circle,
        draw,
        minimum size=1.05cm,
        inner sep=0pt,
        font=\sffamily\Large\bfseries
    },
    cycle arrow/.style={
        very thick,
        -{Stealth[length=3mm,width=2.2mm]}
    }
]

\node[main node] (A1) at (90:2cm)  {$\textcolor{red}{c^{(1)}}$};
\node[main node] (A3) at (-30:2cm) {$\textcolor{red}{c^{(2)}}$};
\node[main node] (A2) at (210:2cm) {$\textcolor{red}{c^{(3)}}$};

\draw[cycle arrow]
    (75:2cm)
    arc[start angle=75,end angle=-15,radius=2cm]
    node[midway,right=2pt] {$\succ$};

\draw[cycle arrow]
    (-45:2cm)
    arc[start angle=-45,end angle=-135,radius=2cm]
    node[midway,below=2pt] {$\succ$};

\draw[cycle arrow]
    (195:2cm)
    arc[start angle=195,end angle=105,radius=2cm]
    node[midway,left=2pt] {$\succ$};
\end{tikzpicture}}
\caption{Illustration of the cyclic clone concept used in \texttt{infMILP}. The abstract alternative type $\textcolor{red}{c}$ is a cycle of clones. While no single clone can form a CWS, the multiset committee $W=\{\textcolor{red}{c}, \textcolor{red}{c}\}$ covers the entire cycle.}
\label{fig:infmilp-cycle-example}
\end{figure}

To construct \infmilp, let $W(c)$ denote the multiplicity of alternative type $c$ in the multiset committee $W$, and let $\supp(W)=\{c\in C \mid W(c)>0\}$ denote the set of alternative types in $W$. We define $P(c,W)$ as the support of the strongest type $c$ against $W$, \emph{i.e.},
\[    
P(c, W) = \begin{cases}
\displaystyle{\sum_{\substack{s\in\cS\\ c\succ_s \supp(W)}}} \vx[s], & \text{if } W(c) = 0 \\
\frac{1}{W(c)} \smashoperator{\sum_{\substack{s\in\cS\\c\succ_s \supp(W)\setminus\{c\}}}}
\vx[s], & \text{if } W(c) > 0.
    \end{cases}    
\]

When $W(c)=0$, the challenger is an alternative type not in the committee, and $P(c,W)$ is simply the fraction of voters who rank $c$ above every type in $W$. When $W(c)>0$, the challenger is an unselected clone of an alternative type that appears in the committee. If the committee contains $W(c)$ clones of type $c$, these clones (in the worst case) can be spaced approximately evenly around the cycle. As the number of clones tends to infinity, the strongest remaining clone defeats all selected clones of type $c$ on a $1/W(c)$ fraction of the cycle. 

Using this formulation, we define \infmilp~ as
\begin{align*}
    \max \quad & \alpha \\
    \text{s.t.} \quad & \sum_{s\in\cS} \vx[s] = 1  \\
    & \sum_{c\in C}\vy_{W,c} = 1, \quad \forall W\in \cM^+_k \\
    &  P(c, W) + (1-\vy_{W,c}) \ge \alpha, \quad \forall W\in \cM^+_k, c\in C \\
    & \vx \ge 0, \quad \vy_{W,c} \in \{0, 1\}, \quad \alpha \ge 0 
\end{align*}

Unlike \milp, the optimal value of \infmilp~describes an infinite alternative election. An objective value strictly greater than $\alpha$ guarantees a finite counterexample that no $\alpha$-undominated set exists.

\begin{restatable}{theorem}{infMILPCorrectness}
\label{thm:infmilp-correctness}
Fix $m$ and $k$, and let $\alpha^*$ be the optimal value of \infmilp. Then, for every
$\alpha<\alpha^*$, there exists a finite election with strict ordinal preferences for which no $\alpha$-undominated set of size $k$ exists.
\end{restatable}

Compared to \milp, \infmilp~ offers greater modeling power at the cost of increased complexity. While the number of possible committees grows to $\binom{m+k-1}{k}$, the dominant factor in the program's size is still the $m!$ rankings. In~\Cref{sec:appendix-bm}, we introduce yet another equivalent formulation based on \textit{Block--Marschak} choice probabilities~\citep{Block59:Random,Falmagne78:Representation}, reducing the factorial representation to an exponential size in $m$. As discussed in~\Cref{sec:experiments}, this reformulation substantially reduces memory requirements and enables searches at larger $m$. However, for smaller instances, the original method is computationally faster, highlighting a tradeoff between memory efficiency and solver performance. 

\subsection{Theoretical lower bounds}

By constructing feasible solutions to our programs, we can immediately obtain lower bounds on $\alpha^*$ that hold across all parameter values. For \milp, instantiating $\vx$ as the uniform distribution over rankings gives the following bound.   

\begin{restatable}{proposition}{MILPLowerBound}[Lower bound for \milp]
\label{prop:milp-lower-bound}
Let $\alpha^*$ be the optimal value of \milp. For every $m > k$,
$$ \alpha^* \geq \frac{1}{k+1}. $$
When $m = k+1$, $\alpha^* = \frac{1}{k+1}$.
\end{restatable}

When $m=k+1$, the exact value of \milp~ is only $\nicefrac{1}{(k+1)}$, which is below the known general lower bound of $\nicefrac{2}{(k+1)}$. This general lower bound permits arbitrarily many alternatives, whereas any parameter assignment of \milp~ fixes their number. This distinction motivates the clone construction of \infmilp.  In particular, the known general lower bound can be recovered from \infmilp.

\begin{restatable}{proposition}{InfMILPLowerBound}[Lower bound for \infmilp]
\label{prop:infmilp-lower-bound}
Let $\alpha^*$ be the optimal value of \infmilp. For every $m > k$,
$$ \alpha^* \geq \frac{2}{k+1}. $$
\end{restatable}

The instantiation of \infmilp~assigns equal probability to the $k+1$ cyclic rotations of alternative types $1,\ldots,k+1$, with all remaining types ranked below them in a fixed order. This lower bound for \infmilp~is useful when searching for counterexamples as an incumbent solution always exists at the edge of known bounds.

\subsection{Theoretical upper bound}

As discussed in \Cref{sec:intro}, prior theoretical approaches were able to produce upper bounds for $\alpha^*$ in terms of $k$; these include $\alpha \leq 1 - \frac{k-1}{k^{k/(k-1)}}$ and $\alpha \leq \frac{4.910...}{k+1}$ \citep{Charikar26:Exposition,Song26:Few}. While these upper bounds hold for all values of $m$ (and are thus very general), they still leave a gap with the conjectured $\alpha=\nicefrac{2}{(k+1)}$ value. In this section, we use our framework to show that by tolerating upper bounds that depend on both $m$ and $k$, we can achieve tighter bounds in wide regimes. In fact, as we will see, our bound can be used to show that the conjectured value of $\nicefrac{2}{(k+1)}$ can be achieved (and in fact surpassed) whenever $k$ is larger than a constant fraction of $m$.

\begin{restatable}{theorem}{MILPUpperBound}[Upper bound for \milp]
\label{thm:milp-upper-bound}
Let $\alpha^*$ be the optimal value of \milp. For every $m > k$,
$$ \alpha^* \leq \min_{t \in \mathbb{Z}, 0 \leq t \leq k}\frac{m-t}{(t+1)(k-t+1)}. $$
\end{restatable}
\begin{proof}
   Fix some $t \in \{0,1,\ldots,k\}$ and let $\cM_t = \{W \subseteq C \mid |W| = t\}$ be all subsets of alternatives of size $t$. For a fixed ranking $s \in \cS$, 
   \begin{align}
       \sum_{R \in \cM_t}\sum_{a \in C \setminus R} \mathbbm{1}[a \succ_s R] = {m \choose {t+1}}, \label{eq:count_eq}
   \end{align}
   since $\cM_{t+1}$ can be bijectively mapped to pairs $(R,a)$ with $a \succ_s R$: for any such set in $\cM_{t+1}$, let $a$ be its element ranked highest by $s$, and $R$ the remaining elements. Sampling $s$ from a voter profile $\vx$ preserves the equality in \eqref{eq:count_eq}. By an averaging argument, there then exists some $R \in \cM_t$ with 
    \begin{align}
       \sum_{s \in \cS} \vx[s] \sum_{a \in C \setminus R} \mathbbm{1}[a \succ_s R]   \leq \frac{ {m \choose {t+1}} }{ {m \choose {t}}}=\frac{m-t}{t+1}.\label{eq:count_ineq}
   \end{align}   
   Let \(B(R)= \left\{a \notin R \mid \sum_{s \in \cS \mid a \succ_s R} \vx[s] > \frac{m-t}{(t+1)(k-t+1)}\right\}\). We must have $|B(R)| \leq k-t$, as $|B(R)| \geq k-t+1$ would violate \eqref{eq:count_ineq}. Thus, $R \cup B(R)$ contains at most $k$ alternatives and we can choose some $W \in \cM_k$ such that $R \cup B(R) \subseteq W$. Then for any $a \notin W$, we have
   \[
   \sum_{s \in \cS \mid a\succ_s W}\vx[s] \leq  \sum_{s \in \cS \mid a\succ_s R}\vx[s] \leq \frac{m-t}{(t+1)(k-t+1)},
   \]
   since $R \subseteq W$ and $a \notin B(R)$. Since this is true for all voter profiles $\vx$ and any $0 \leq t \leq k$, we get the theorem statement.
\end{proof}

By setting $t$ to be a constant fraction of $k$, we can recover upper bounds that depend only on $k$, conditional on the committee size being a sufficiently large fraction of the alternative pool.

\begin{corollary}\label{cor:linear_regime}
    For any $c >1$, if $k+1 \geq \frac{4c}{(c+1)^2}m$, then $\alpha^* < \frac{c}{k+1}$.
\end{corollary}
\begin{proof}
    Let $\beta = \frac{c+1}{2c}(k+1)$ and $t= \left\lceil \beta \right\rceil-1$. Since $c>1$, we have $t \leq k$, so \Cref{thm:milp-upper-bound} applies.  Hence, it is sufficient to show 
    \(
       \frac{(m-t)}{(t+1)(k-t+1)}<\frac{c}{k+1} \), or equivalently $f(t) \coloneq t + \frac{c(t+1)(k-t+1)}{k+1} >m$. Since $f(t)$ is concave in $t$ with $f(\beta-1)=f(\beta)$ and $t \in [\beta-1,\beta)$,
       \[
       f(t) \geq f(\beta)= \frac{(c+1)^2(k+1)}{4c}+\frac{c-1}{2} > \frac{(c+1)^2(k+1)}{4c} \geq m,
       \]
       completing the proof.
\end{proof}

By setting $c=2$, \Cref{cor:linear_regime} shows that the conjectured upper bound of $\alpha=\frac{2}{k+1}$ is in fact achieved whenever $k+1 \geq (\nicefrac{8}{9})m$. To the best of our knowledge, this is the first result proving the existence of a linear regime in which the conjecture holds. Similarly, our bound improves on the prior upper bound of $\frac{4.910...}{k}$ whenever $k+1 \geq 0.563m$. In the language of \emph{stable committees} (\Cref{sec:related}), \Cref{cor:linear_regime} proves that a $c$-stable committee exists (even when using the Droop quota) whenever $k +1 \geq \frac{4c}{(c+1)^2}m$. 

Having completed our theoretical analysis, we next return to our experimental results using \milp~ and \infmilp~.
\section{Computational bounds on stability}
\label{sec:experiments}

\ifthenelse{\boolean{isSingleColumn}}
{\renewcommand{\tabwidth}{\linewidth}}
{\renewcommand{\tabwidth}{0.9\linewidth}}

\begin{table*}[t!]
    \centering
    \fontsize{8pt}{11pt}\selectfont
    \caption{Max $\alpha$ values from $\texttt{MILP}$ with an 81,400s timeout. The final row gives the reference threshold $\nicefrac{2}{(k+1)}$.}
    \label{tab:milp-results}
    \begin{tabularx}{\tabwidth}{c|Y|Y|Y|Y|Y|Y|Y}
    \toprule
        \textbf{m} \textbackslash ~\textbf{k} & \textbf{1} & \textbf{2} & \textbf{3} & \textbf{4} & \textbf{5} & \textbf{6} & \textbf{7} \\
        \midrule
        \textbf{2} & 0.500 & -- & --  &  -- & -- &  -- & -- \\
        \textbf{3} & 0.667 & 0.333 & --  &  -- & -- &  -- & -- \\
        \textbf{4} & 0.750 & 0.400 & 0.250  & -- & --  &  -- & --\\
        \textbf{5} & 0.800 & 0.467 & 0.300 &  0.200 & --  &  -- & --\\
        \textbf{6} & 0.833 & 0.500 &  0.333  & 0.231 & 0.167  &  -- & --\\
        \textbf{7} & 0.857 & 0.524 & [0.353, 0.400] &  0.250 & 0.190  &  0.143 & --\\
        \textbf{8} & 0.875 & [0.532, 0.600] & [0.375, 0.444] & [0.275, 0.333] & [0.203, 0.250]  & 0.160 & 0.125 \\
       \midrule
       $\mathbf{\nicefrac{\mathbf{2}}{(k+1)}}$ & 1.000 & 0.667 & 0.500 & 0.400 & 0.333 & 0.286 & 0.250 \\ 
       \bottomrule
    \end{tabularx}
\end{table*}

We implement \milp~ and \infmilp~using the Gurobi Optimizer version 12.0.3~\citep{Gurobi26:Gurobi} with a tolerance of $10^{-6}$. Experiments are conducted on a high-performance computing cluster. Each node is equipped with dual-socket AMD EPYC 7282 processors (32 physical cores, 64 logical threads) and 512 GB of RAM, running Linux kernel 4.18. 

We report in~\Cref{tab:milp-results} the results of running \milp~ across values of $m$ and $k$. A single value indicates that the instance was solved to optimality. An interval $[a,b]$ indicates the solver timed out with a feasible solution $\alpha=a$ and proved an upper bound $\alpha \le b$. The results from \milp~ prove that \textit{all elections} with $m \le 8$ alternatives and $3 \le k < m$ admit a CWS, independent of the number of voters. More generally, every upper bound is strictly below $\nicefrac{2}{(k+1)}$, proving the existence of a $\nicefrac{2}{(k+1)}$-undominated committee for every election with at most eight alternatives. The outputs of \milp~also certify the worst-case value of $\alpha$ which defines the threshold above which an $\alpha$-undominated committee is guaranteed to exist, providing insight into elections under these parameter assignments. 

An observation from~\Cref{tab:milp-results} is that the value of $\alpha^*$ grows with $m$ for a fixed $k$. Nondecreasing behavior is expected: adding alternatives ranked last by every voter cannot decrease the worst-case value. However, these values are still far from the conjectured upper bound. \Cref{thm:milp-upper-bound} above complements these computational results with a theoretical upper bound in a linear regime relating $k$ and $m$.

While \milp~certifies that no counterexamples exist for $m \le 8$, it is still possible that \infmilp~ constructs a counterexample in this parameter regime as it is not restricted to finite alternatives. We include a subset of values of \infmilp~in \Cref{tab:infmilp-results} (the full table can be found in~\Cref{sec:omitted-tables}). We see the advantage of \infmilp~through the tight lower bound on $\alpha^*$ for all configurations; even at small parameter values, \infmilp~computes solutions at the edge of known bounds. Our results for \infmilp~certify that no counterexamples exist for $m \le 5$ alternative types (recall that in \infmilp, each $c\in C$ represents infinitely many clones) and for $m = 6$ and $k \le 3$, proving the known lower bound is tight at these values. 

\ifthenelse{\boolean{isSingleColumn}}
{\renewcommand{\tabwidth}{0.6\linewidth}}
{\renewcommand{\tabwidth}{0.9\linewidth}}

\begin{table}[b!]
    \centering
    \caption{Max $\alpha$ values from $\texttt{infMILP}$ with an 81,400s timeout. The final row gives the reference threshold $\nicefrac{2}{(k+1)}$.}
    \label{tab:infmilp-results}
    \begin{tabularx}{\tabwidth}{c|Y|Y|Y}
        \toprule
        \textbf{m} \textbackslash ~\textbf{k} &\textbf{2} & \textbf{3} & \textbf{4} \\
         \midrule
       \textbf{3} & 0.667 & --  &  --  \\
       \textbf{4}   &  0.667 & 0.500 & -- \\
       \textbf{5}  & 0.667 & 0.500 &  0.400 \\
       \textbf{6}   & 0.667 &  0.500  & [0.400, 0.500]  \\
       \textbf{7}   & 0.667 & 0.500 & [0.400, 0.500]  \\
       \textbf{8}   & [0.667, 0.750] & [0.500, 0.667] & [0.400, 0.667] \\
        \textbf{9} & [0.667, 0.800]  & [0.500, 0.750] & [0.400, 0.750] \\
        \midrule
       $\mathbf{\nicefrac{\mathbf{2}}{(k+1)}}$  & 0.667 & 0.500 & 0.400 \\ 
       \bottomrule
    \end{tabularx}
\end{table}

For larger values of $m$, the runs did not terminate with an exact solution, and therefore we cannot rule out the existence of a counterexample. We target these parameter regimes with more pointed searches; we repeatedly randomly sample a large (but not exhaustive) number of rankings $\cS' \subset \cS$ and run \infmilp~to termination. For the parameter settings in which we did not obtain an exact solution for $7 \le m \le 9$, performing this additional search did not yield any counterexamples.

The main limitation of \milp~and \infmilp~is scalability. Despite being independent of $n$, the ranking formulations have factorial dependence on $m$, while the equivalent Block--Marschak formulations have exponential dependence. Under our computational budget, we did not obtain informative bounds beyond
$m=8$ for \milp~and $m=9$ for \infmilp. 

The smaller representation also did not uniformly improve computational performance. At smaller values of $m$, the explicit ranking formulations were faster in our experiments, with Gurobi processing branch-and-bound nodes more quickly. The principal advantage of the Block--Marschak formulations was their lower memory requirements: they enabled runs at $m\ge8$ that were otherwise limited by the available memory.

\section{Beyond stability: connecting other axioms}
\label{sec:axioms}

Stability is just one property that we can use to measure the quality of a collective decision. We next study the implications between stability and other notions of proportionality for ranked elections.

Rank-priceability~\citep{Brill23:Robust} is an extension of priceability from the approval-based setting~\citep{Peters20:Proportionality} to ranked preferences. In this setup, each voter is given an equal budget, and every elected alternative must be purchased at the same price. For every unelected alternative and rank cutoff $r$, consider the voters who rank that alternative among their top $r$ alternatives. Their collective unspent budget, together with their payments to elected alternatives outside their top $r$, must not exceed the common price.

\begin{definition}[Rank-priceability]
Let $W \in \cM_k$. For a ranking $s_i$, let $\operatorname{top}_r(s_i)$ denote voter $i$'s top-$r$ alternatives. The committee $W$ is \emph{rank-priceable} if there exist payments $f_{i,w}\geq 0$ for every voter $i\in N$ and committee member $w\in W$, and a common price $\tau > 0$, so that:
\begin{enumerate}
    \item $\forall i\in N$, $\sum_{w\in W} f_{i,w}\leq\frac{1}{n},$

    \item $\forall w\in W$, $\sum_{i\in N} f_{i,w}=\tau,$ and

    \item $\forall c\notin W$ and $\forall r\in[m]$,
    $$
    \sum_{i\in N \mid c\in\operatorname{top}_r(s_i)}
    \left( \frac{1}{n} - \sum_{w\in W\cap\operatorname{top}_r(s_i)} f_{i,w} \right)  \leq \tau.
    $$
\end{enumerate}
\end{definition}

\Cref{lem:priceability-lp} gives a linear characterization of rank-priceability whose size is independent of the number of voters. For a fixed committee $W$, the constraints are linear in the profile $\vx$, the aggregate payments $\vf$, and the common price $\tau$. For a given $m$, $k$, $\vx$, and $W$, we can determine if $W$ is rank-priceable via a linear program.

\begin{restatable}{lemma}{PriceMILP}
\label{lem:priceability-lp}
Fix $m$, $k$, and $n$, and let $\vx\in\Delta(\cS)$ denote the distribution of rankings in an election with $n$ voters. A committee $W\in\cM_k$ is rank-priceable if and only if there exist nonnegative  payments $\vf=\{\vf[s,w]\}_{s\in\cS,w\in W}$ and a common price $\tau > 0$
satisfying:
\begin{enumerate}
    \item $\forall s\in\cS$,  $\sum_{w\in W}\vf[s,w]\leq\vx[s]$,
    \item $\forall w\in W$, $\sum_{s\in\cS}\vf[s,w]=\tau$, and 
    \item $\forall c\notin W$ and $r\in[m]$,
    $$
    \sum_{s\in\cS \mid c\in\operatorname{top}_r(s)}
    \left(\vx[s] - \sum_{w\in W\cap\operatorname{top}_r(s)}\vf[s,w]\right)
    \leq\tau.
    $$
\end{enumerate}
\end{restatable}

We focus on \textit{incomparability}; we show that rank-priceability need not imply stability and, on the contrary, that stability need not imply priceability. Here, stability refers to the committee under consideration being $2/(k+1)$-undominated. For each incomparability result we prove in this section, we present the \emph{minimal} $(m,k)$ for which the implication breaks, meaning the implication holds true for all $(m',k')$ such that $m'<m$ or $m'=m$ and $k'<k$.

To find a rank-priceable committee that is not $\alpha$-undominated, we impose the payment constraints and maximize an unelected alternative's support against $W$. This is a linear program. For the reverse direction, we use the dual of the linear program that determines whether $W$ is rank-priceable. A feasible dual solution with strictly positive objective certifies that $W$ is not rank-priceable. We then minimize the largest unelected alternative's support subject to this certificate. The resulting program is bilinear. We give both formulations in~\Cref{sec:appendix-axioms}.

In fact, the approach to prove incomparability is not unique to rank-priceability. In~\Cref{sec:appendix-axioms}, we give a general approach to incorporate axioms defined by linear constraints into \milp~ to measure the implications between axioms. 

The programs produce voter profiles and payment or nonpriceability certificates from which we obtain the following incomparability results. 


\begin{restatable}{theorem}{PriceabilityStability}[Rank-priceability $\not\Rightarrow$ stability]
\label{thm:price-stability}
    There exists a voter profile for $m=2$ and $k=1$ such that there is a $W \in \cM_k$ that satisfies rank-priceability, but is not $\alpha$-undominated for $\alpha = 1$.
\end{restatable}

Despite being the smallest example showing separation between rank-priceability and $\nicefrac{2}{(k+1)}$- undominatedness, this counterexample is degenerate since $k=1$ and $\alpha = 1$. We note that the smallest non-degenerate counterexample (with $k >1$ and $\alpha<1$) exists for $m=4$ and $k=2$. We also disprove the implication in the other direction using an analogous program. 

\begin{restatable}{theorem}{StabilityPriceability}[Stability $\not\Rightarrow$ rank-priceability]
\label{thm:stability-price}
    There exists a voter profile for $m=3$ and $k=2$ such that there is a $W \in \cM_k$ that is $\alpha$-undominated for $\alpha = \nicefrac{2}{3}$, but not rank-priceable.
\end{restatable}

To illustrate just how flexible this approach is (and thus the generality of our framework), we obtain incomparability results (in both directions) for two additional axioms: Rank-PJR+ and Rank-EJR+~\citep{Brill23:Robust}.  Unlike rank-priceability, the incomparability programs for these axioms remain linear. We include the results for Rank-PJR+ here, and the results for Rank-EJR+ can be found in~\Cref{sec:appendix-axioms}.

\begin{definition}[Rank-PJR+]
    Let $W\in\cM_k$. For a ranking $s_i$, let $\operatorname{top}_r(s_i)$ denote voter $i$'s top-$r$ alternatives. The committee $W$ satisfies Rank-PJR+ if, for all $r\in[m]$, all $c\in C \setminus W$, all $\ell\in[k]$, and all voter subsets $N'\subseteq N$ with
    $$
    |N'|\geq \frac{\ell}{k}n,
    $$
    if $c\in\operatorname{top}_r(s_i)$ for all $i \in N'$ then
    $$
    \left| W\cap\bigcup_{i\in N'}\operatorname{top}_r(s_i) \right| \geq \ell.
    $$
\end{definition}

As before, we can represent this axiom within our framework via linear constraints that are independent of $n$.

\begin{restatable}{lemma}{PJRMILP}
\label{lem:pjrp-milp}
Fix $m$, $k$, and $n$, and let $\vx\in\Delta(\cS)$ denote the distribution of rankings in an election with $n$ voters. A committee $W\in\cM_k$ satisfies Rank-PJR+ if and only if, for every $c\notin W$, $r\in[m]$, and proper subset $T \subsetneq W$,
$$
\sum_{\substack{s\in\cS \mid c\in\operatorname{top}_r(s), \\ W\cap\operatorname{top}_r(s)\subseteq T}}
\vx[s] < \frac{|T|+1}{k}.
$$
\end{restatable}

Using this characterization, we obtain the following result.  

\begin{restatable}{theorem}{PJRStability}[Rank-PJR+ $\not\Rightarrow$ stability]
\label{thm:pjr-stability}
    There exists a voter profile for $m=3$ and $k=1$ such that there is a
    $W\in\cM_k$ that satisfies Rank-PJR+, but is not $\alpha$-undominated for $\alpha=1$.
\end{restatable}

And likewise, we obtain an incomparability result in the opposite direction through another program. 

\begin{restatable}{theorem}{StabilityPJR}[Stability $\not\Rightarrow$ Rank-PJR+]
\label{thm:stability-pjr}
    There exists a voter profile for $m=3$ and $k=2$ such that there is a
    $W\in\cM_k$ that is $\alpha$-undominated for $\alpha=\nicefrac{2}{3}$, but does not  satisfy Rank-PJR+.
\end{restatable}

The preceding results concern individual committees: a committee can satisfy one property while violating the other. A stronger question concerns \emph{incompatibility}: are there elections with no committee that is simultaneously $\alpha$-undominated for $\alpha=\nicefrac{2}{(k+1)}$ and satisfies an additional axiom?

Our framework can also address this question. For every committee, we can enforce that there exists an unelected alternative with support at least $\alpha$ or that there exists a certificate that the committee violates the additional axiom. For Rank-PJR+, both certificates are expressed by linear inequalities, so the resulting incompatibility program is a MILP.

\Cref{tab:rank-pjrp-results} reports this program's optimal objective values for $m\leq6$. Every value is strictly below $\nicefrac{2}{(k+1)}$, meaning that for any number of voters, there exists a committee that satisfies Rank-PJR+ \emph{and} is $\nicefrac{2}{(k+1)}$-undominated. While we proved incomparability among these properties, they are always simultaneously achievable in elections with at most six alternatives. In fact, in this parameter regime, satisfying Rank-PJR+ results in the exact same values of $\alpha^*$ as \milp. This result raises a natural question: can Rank-PJR+ always be satisfied without hurting the worst-case stability guarantee?

\begin{table}[b!]
    \centering
    \caption{Max $\alpha$ values from $\texttt{MILP}$ with incompatibility constraints for Rank-PJR+. The final row gives the reference threshold $\nicefrac{2}{(k+1)}$.}
    \label{tab:rank-pjrp-results}
    \begin{tabular}{c|c|c|c|c|c}
    \toprule
        \textbf{m} \textbackslash ~\textbf{k} & \textbf{1} & \textbf{2} & \textbf{3} & \textbf{4} & \textbf{5}\\
        \midrule
        \textbf{2} & 0.500 & -- & --  &  -- & -- \\
       \textbf{3} & 0.667 & 0.333 & --  &  -- & -- \\
       \textbf{4} & 0.750 & 0.400 & 0.250  & -- & -- \\
       \textbf{5} & 0.800 & 0.467 & 0.300 & 0.200 & -- \\
       \textbf{6} & 0.833 & 0.500 & 0.333 & 0.231 & 0.167 \\
       \midrule
       $\mathbf{\nicefrac{\mathbf{2}}{(k+1)}}$ & 1.00 & 0.667 & 0.500 & 0.400 & 0.333 \\ 
       \bottomrule
    \end{tabular}
\end{table}
\section{Conclusions}

We developed a framework leveraging automated reasoning to study the worst-case stability of a committee in ranked choice voting. Our formulation computes bounds that are independent of the number of voters and generalize to all elections for a given number of alternatives and committee size. We proved computationally that all elections with at most eight alternatives admit a $\nicefrac{2}{(k+1)}$-undominated committee. In addition to the computational results, we established theoretical lower bounds for our programs and an upper bound that proves that a $\nicefrac{2}{(k+1)}$-undominated committee exists whenever $k\geq \nicefrac{8m}{9}$. We also extended our programs to various proportionality axioms, obtaining the smallest sized incomparability examples, and investigating whether
stability and proportionality can always be achieved simultaneously.

Our results open several additional paths for future research. While we gave a regime for which a $\nicefrac{2}{(k+1)}$-undominated committee exists, it is still elusive whether this holds universally, and in particular if a CWS of size three or four always exists. Scaling our formulations could reveal new counterexamples or establish existence guarantees over larger parameter ranges. The resulting constructions and bounds may also expose structural properties that lead to more general proofs.

The relationship between committee stability and the proportionality axioms studied here raises further open questions. Although we established incomparability, imposing Rank-PJR+ does not change the optimal stability guarantee for the parameters tested. Can Rank-PJR+ always be satisfied without sacrificing the optimal worst-case stability guarantee? More broadly, which additional axioms do not change the optimal worst-case stability guarantee? Answering these questions would shed light on the relationship between proportional representation and committee stability. 

Overall, our methods provide a systematic way to explore these questions through both counterexample searches and existence certificates. Together with related automated reasoning approaches in approval-based committee selection~\citep{Berker25:Edge}, our results demonstrate the value of optimization-based methods for studying collective choice. We anticipate and hope that our techniques will be fruitful for tackling additional problems in social choice. 
\section*{Acknowledgments}

The authors would like to thank Prasanna Ramakrishnan and Ioannis Anagnostides for helpful discussions (in particular, on \infmilp~and \Cref{thm:milp-upper-bound}, respectively). I.Z.\ is supported by NIH award A240108S001, the Vannevar Bush Faculty Fellowship ONR N00014-23-1-2876, and National Science Foundation grant RI-2312342, as well as the NSF Graduate Research Fellowship Program under grants DGE2140739 \& DGE2631988. R.E.B.\ thanks the Cooperative AI Foundation, Macroscopic Ventures (formerly Polaris Ventures / the Center for Emerging Risk Research), and Jaan Tallinn’s donor-advised fund at Founders Pledge for financial support. R.E.B.\ is also supported by the Cooperative AI PhD Fellowship.  Any opinions, findings, and conclusions or recommendations expressed in this material are those of the author(s) and do not necessarily reflect the views of the funding agencies.

\bibliography{references}

\clearpage

\appendix
\section{Omitted roofs from \Cref{sec:milps}}
\label{sec:ommitted-proofs}

\subsection{Proof of \Cref{prop:opt-correctness}}

\OptCorrectness*

\begin{proof}
($\Leftarrow$) Suppose first that $\alpha>\alpha^*$. For any voter profile $\vx$, by the definition of $\alpha^*$,
\[
\min_{W\in\cM_k}\max_{c\notin W} \smashoperator{\sum_{\csuccW}}\vx[s] \leq \alpha^* <\alpha.
\]
Therefore, there exists some $W\in\cM_k$ such that
\[ \max_{c\notin W} \smashoperator{\sum_{\csuccW}}\vx[s] <\alpha. \]
So $W$ is an $\alpha$-undominated set.

($\Rightarrow$) Suppose on the contrary that $\alpha\leq\alpha^*$, and let $\vx^*$ be a profile attaining $\alpha^*$ in~\ref{eq:alphastar}. Then
\[ \min_{W\in\cM_k}\max_{c\notin W} \smashoperator{\sum_{\csuccW}}\vx^*[s] =\alpha^*. \]
For every $W\in\cM_k$, there exists some $c\notin W$ such that
\[
\smashoperator{\sum_{\csuccW}}\vx^*[s] \geq \alpha^* \geq \alpha.
\]
Therefore, no committee $W\in\cM_k$ is $\alpha$-undominated.
\end{proof}

\subsection{Proof of \Cref{thm:milp-correctness}}

\MILPCorrectness*

\begin{proof}
    Let $(\vx, \vy, \alpha)$ be any solution to \milp. For each $W$, the constraints enforce that there is one $c_W \notin W$ with $\vy[W,c_W] = 1$. The corresponding constraint on $\alpha$ ensures that 
    $$\alpha \le \smashoperator{\sum_{s \in \cS \mid c_W \succ_s W}} \vx[s] \le \max_{c \notin W} \smashoperator{\sum_{\csuccW}} \vx[s]. $$

    Taking the minimum over $W \in \cM_k$, we obtain

    $$ \alpha \le \min_{W\in \cM_k} \max_{c \notin W} \smashoperator{\sum_{\csuccW}} \vx[s] \le \alpha_1^*.$$

    The last inequality follows from $\vx \in \Delta(\cS)$. Since this holds for every feasible solution of \milp, we have $\alpha_2^* \le \alpha_1^*$. 

    To prove the reverse inequality, let $\vx^*$ be any optimal profile of~\ref{eq:alphastar}. For each $W\in \cM_k$, choose 

    $$ c_W \in \arg\max_{c \notin W} \smashoperator{\sum_{\csuccW}} \vx^*[s], $$

    and set $\vy[W,c_W] = 1$ and $\vy[W,c] = 0$ for $c \notin W$ with $c \neq c_W$. We show that $(\vx^*, \vy, \alpha_1^*)$ is feasible for \milp. First, the constraints $\sum_{c\notin W} \vy[W,c] = 1$ and $\vy[W,c] \in \{0,1\}$ hold by construction for all $W \in \cM_k$. Since $\vx^* \in \Delta(\cS)$, it also follows that $\sum_{s \in \cS} \vx^*[s] = 1$. What remains to verify is the constraint on $\alpha$. For each $c_W$, optimality of $\vx^*$ implies 

    $$ \smashoperator{\sum_{s \in \cS \mid c_W \succ_s W}} \vx^*[s] = \max_{c \notin W} \smashoperator{\sum_{\csuccW}} \vx^*[s] \ge \alpha_1^*. $$

    For $c \notin W$ with $c \neq c_W$, the constraint relaxes to $Q \ge \alpha_1^*$ since the summation is nonnegative. This is always satisfied for $Q\ge 1$. So $(\vx^*, \vy, \alpha_1^*)$ satisfies all constraints of \milp, and is therefore feasible. It follows that $\alpha_1^* \le \alpha_2^*$.

    Since $\vx^*$ is an arbitrary optimal profile for~\ref{eq:alphastar}, this construction also shows that every such profile extends to an optimal solution of \milp. The first part of the proof already gives the reverse implication, so every $\vx$ that is optimal for \milp~is also optimal for \ref{eq:alphastar}.

\end{proof}

\subsection{Proof of \Cref{prop:symmetry-milp}}

\MILPSymmetryCorrectness*

\begin{proof}
Consider any feasible solution of the symmetry broken \milp. For every $W\neq W^*$, the constraints imply
\[ \alpha \leq \max_{c\notin W} \smashoperator{\sum_{\csuccW}}\vx[s]. \]
For $W^*$, we have
\[ \alpha = \smashoperator{\sum_{s\mid c_{k+1}\succ_s W^*}}\vx[s] \leq \max_{c\notin W^*} \smashoperator{\sum_{s\mid c\succ_s W^*}}\vx[s]. \]
Therefore,
\[ \alpha \leq \min_{W\in\cM_k} \max_{c\notin W} \smashoperator{\sum_{s\mid c\succ_s W}}\vx[s] \leq \alpha^*. \]
So the optimal value of the symmetry broken \milp~ is at most $\alpha^*$.

Conversely, take an optimal profile $\vx^*$ for~\ref{eq:alphastar} with value $\alpha^*$. Let $W$ be the minimizing committee and $c\notin W$ its challenger. Relabel alternatives so that $W=\{c_1,\ldots,c_k\}$ and  $c=c_{k+1}$. Alternative relabeling preserves all probabilities and the objective value. So the symmetry broken \milp~ has a feasible solution with value $\alpha^*$. Therefore, the optimal value is $\alpha^*$.
\end{proof}

\subsection{Proof of \Cref{thm:infmilp-correctness}}

\infMILPCorrectness*

\begin{proof}
Let $\vx^*$ be the optimal profile for \infmilp~ and let $\epsilon=\alpha^*-\alpha>0$. Choose an integer $K>k$ sufficiently large such that $\frac{1}{K}<\epsilon$, and replace every alternative type by $K$ cyclically ranked clones.

Consider any committee of $k$ clones, and let $W\in\cM_k^+$ be the corresponding multiset. By the definition of $\alpha^*$, there exists some alternative type $c$ such that
\[
P(c,W)\geq\alpha^*.
\]
If $W(c)=0$, an unselected clone of type $c$ has challenger support exactly
$P(c,W)$. If $W(c)=r>0$, the $r$ selected clones partition the $K$-cycle
into $r$ segments. The largest segment has size at least $\lceil K/r\rceil$, so
there exists an unselected clone of type $c$ that defeats all $r$ selected
clones on at least $\lceil K/r\rceil-1$ of the $K$ cyclic rotations.
Therefore, its support is at least
\[
P(c,W)-\frac{1}{K}.
\]
So, every committee of size $k$ has an outside challenger supported by at least
\[
\alpha^*-\frac{1}{K} > \alpha
\]
of the electorate. We conclude that no committee is $\alpha$-undominated.
\end{proof}

\subsection{Proof of \Cref{prop:milp-lower-bound}}

\MILPLowerBound*

\begin{proof}
By the equivalence of \milp~ and the max-min formulation, it suffices to construct a voter profile attaining value $1/(k+1)$. This can be achieved via the uniform distribution over all $m!$ strict rankings of the alternatives, 
$$
\vx[s]=\frac{1}{m!}, ~\forall s\in\cS.
$$
Fix any committee $W\in\cM_k$ and any alternative $c\notin W$. Alternative $c$ defeats the committee exactly when it is ranked above every member of $W$. Among the $k+1$ alternatives in $W\cup\{c\}$, each alternative is equally likely to be ranked highest under the uniform distribution. Therefore,
$$ \sum_{\substack{s\in\cS\\c\succ_s W}}\vx[s] = \frac{1}{k+1}. $$
Since this holds for every $W\in\cM_k$ and every $c\notin W$,
$$
\max_{c\notin W} \sum_{\substack{s\in\cS\\c\succ_s W}}\vx[s] = \frac{1}{k+1}.
$$
Consequently,
$$
\min_{W\in\cM_k} \max_{c\notin W} \sum_{\substack{s\in\cS\\c\succ_s W}}\vx[s] = \frac{1}{k+1}.
$$
So, $\alpha^*\geq\frac{1}{k+1}$.

To show tightness when $m=k+1$, consider an arbitrary voter profile $\vx$. Every committee $W\in\cM_k$ is obtained by excluding a unique alternative $c$, so $W=C\setminus\{c\}$. Alternative $c$ defeats every member of $W$ exactly when $c$ is ranked first. Let
$$ p_c = \sum_{\substack{s\in\cS\\c\text{ is ranked first in }s}}\vx[s] $$
denote the probability that $c$ is ranked first. Since every ranking has exactly one top-ranked alternative, $\sum_{c\in C} p_c=1$.  There are $k+1$ alternatives, so there exists some $c$ such that $p_c\leq\frac{1}{k+1}$. For the committee $W=C\setminus\{c\}$, $c$ is its only outside challenger, so
$$
\max_{a\notin W} \sum_{\substack{s\in\cS\\a\succ_s W}}\vx[s] = p_c \leq \frac{1}{k+1}.
$$
For every voter profile,
$$
\min_{W\in\cM_k}
\max_{c\notin W}
\sum_{\substack{s\in\cS\\c\succ_s W}}\vx[s]
\leq
\frac{1}{k+1}.
$$
Combining this upper bound with the previous lower bound concludes the proof.
\end{proof}

\subsection{Proof of \Cref{prop:infmilp-lower-bound}}

\InfMILPLowerBound*

\begin{proof}
We construct a profile for which every multiset committee has a challenger with support at least $2/(k+1)$. Selecting one such challenger for each committee gives a feasible \infmilp~ solution. 

Choose $k+1$ distinguished alternative types
$$
C_0=\{0,1,\ldots,k\}\subseteq C.
$$
For every $i\in C_0$, let $s_i$ be a ranking whose restriction to $C_0$ is
$$
i\succ i+1\succ\cdots\succ i+k,
$$
where indices are taken modulo $k+1$. Rank every alternative type in $C\setminus C_0$ below all types in $C_0$, in an arbitrary fixed order. Consider the voter profile
$$
\vx[s_i]=\frac{1}{k+1}, ~\forall i\in C_0,
$$
with zero mass on all other rankings.

Fix any multiset committee $W\in\cM_k^+$. Since $W$ contains $k$ alternatives in total,
$$
\sum_{i\in C_0} W(i) \leq k.
$$

We first show that there exists some $c\in C_0$ such that
$$
W(c-1)=0 \text{ and } W(c) \leq1.
$$
Since there are $k+1$ alternative types in $C_0$ but at most $k$ copies represented in $W$, at least one $W(i)$ is zero. Suppose, for contradiction, that every zero-multiplicity type is followed cyclically by a type of multiplicity at least $2$. Let $z$ be the number of types $i\in C_0$ for which $W(i)=0$. There are $k+1-z$ nonzero types, each with multiplicity at least $1$. Moreover, the $z$ successors of the zero-multiplicity types are distinct and, by assumption, each has multiplicity at least $2$. Therefore,
$$
\sum_{i\in C_0}W(i) \geq (k+1-z)+z = k+1,
$$
contradicting $\sum_{i\in C_0}W(i)\leq k$. So, such a type $c$ exists.

Now consider the two rankings $s_c$ and $s_{c-1}$. In $s_c$, alternative type $c$ is ranked above every other alternative type in $C_0$. In $s_{c-1}$, alternative type $c$ is ranked second, immediately below $c-1$. Since
$$
W(c-1)=0,
$$
alternative type $c-1$ is not represented in $W$. So, in both $s_c$ and $s_{c-1}$, alternative type $c$ is ranked above every represented alternative type in $C_0$ other than itself. By construction, every alternative type outside $C_0$ is ranked below every type in $C_0$, so $c$ is also ranked above every represented type outside $C_0$.

There are now two cases. If $W(c)=0$, then by the definition of $P(c,W)$,
$$
P(c,W) = \sum_{\substack{s\in\cS\\ c\succ_s\supp(W)}} \vx[s].
$$
Both $s_c$ and $s_{c-1}$ appear in this sum, so
$$
P(c,W) \geq \vx[s_c]+\vx[s_{c-1}] = \frac{2}{k+1}.
$$

If $W(c)=1$, then
$$
P(c,W) = \frac{1}{W(c)} \sum_{\substack{s\in\cS\\ c\succ_s\supp(W)\setminus\{c\}}} \vx[s].
$$
Since $W(c)=1$, the multiplicative factor is equal to $1$. Again, both $s_c$ and $s_{c-1}$ appear in the sum, and therefore
$$
P(c,W) \geq \frac{2}{k+1}.
$$

This holds for every multiset committee $W\in\cM_k^+$, so it follows that $\alpha^* \geq \frac{2}{k+1}$.
\end{proof}

\section{A compact reformulation of \infmilp}
\label{sec:appendix-bm}

The primary bottleneck in \infmilp~ is the representation of the voter profile, which contains one variable for each of the $m!$ rankings of the alternative types. However, \infmilp~does not use the complete ranking distribution directly. Instead, it only requires probabilities that a particular alternative type is ranked above a specified set of alternatives. We can exploit this structure to obtain an exact lower-dimensional representation using stochastic choice probabilities.

For every nonempty set $A\subseteq C$ and alternative $c\in A$, define
\[p(c,A) = \Pr_{\vx}\left[c\succ A\setminus\{c\}\right] =
\sum_{\substack{s\in\cS\\ c\succ_s A\setminus\{c\}}} \vx[s].
\]
The value $p(c,A)$ is the probability that $c$ is the highest-ranked alternative among the alternatives in $A$. Since every ranking has exactly one highest-ranked alternative in each nonempty set $A$,
\begin{equation*}
\sum_{c\in A}p(c,A)=1,~
\forall \varnothing\neq A\subseteq C.
\end{equation*}

These normalization constraints alone are insufficient to guarantee that the choice probabilities arise from a common distribution over strict rankings. To enforce this consistency, for every $c\in A\subseteq C$, define the Block--Marschak quantity
\begin{equation*}
\operatorname{BM}(c,A) = \sum_{B \mid A\subseteq B\subseteq C}(-1)^{|B\setminus A|}p(c,B).
\end{equation*}

For unfamiliar readers, we briefly recall the Block--Marschak representation theorem that underlies our reformulation~\citep{Block59:Random,Falmagne78:Representation}. A collection of stochastic choice probabilities $p(c,A)$ specifies, for every nonempty $A\subseteq C$, the probability that each $c\in A$ is chosen as the most-preferred alternative. The Block--Marschak theorem characterizes exactly when such marginal probabilities are jointly consistent with a single probability distribution over strict rankings.

\begin{theorem}[Block--Marschak representation theorem]
\label{thm:block-marschak}
A collection of choice probabilities $\{p(c,A)\mid \varnothing\neq A\subseteq C, c\in A\}$
is induced by a probability distribution over strict rankings of $C$ if and only if
\[ \sum_{c\in A}p(c,A)=1,~\forall \varnothing\neq A\subseteq C 
\text{ and } \operatorname{BM}(c,A)\geq0,~\forall c\in A\subseteq C. \]
\end{theorem}

We can now express every challenger score used by \infmilp~directly in terms of choice probabilities. For a multiset committee $W\in\cM_k^+$ and alternative type $c\in C$, define
\[
A(W,c)=\supp(W)\cup\{c\}
\]
and
\[
\eta(W,c) =
\begin{cases}
1, & W(c)=0,\\
\displaystyle\frac{1}{W(c)}, & W(c)>0.
\end{cases}
\]
Then the two cases in the definition of $P(c,W)$ can be written compactly as
\begin{equation*}
P(c,W) = \eta(W,c)p(c,A(W,c)).
\end{equation*}

Indeed, if $W(c)=0$, then $A(W,c)=\supp(W)\cup\{c\}$, and $p(c,A(W,c))$ is exactly the probability that $c$ is ranked above every type represented in $W$. If $W(c)>0$, then $A(W,c)=\supp(W)$, so the same choice probability gives the fraction of voters who rank $c$ above all other represented types, while the factor $1/W(c)$ accounts for the cycle of clones.

Substituting this into \infmilp~yields the following program.
\begin{align*}
    \max \quad & \alpha \\
    \text{s.t.} \quad & \sum_{c\in A}p(c,A)=1, \quad \forall \varnothing \neq A\subseteq C,\\
    & \operatorname{BM}(c,A)\geq0, \quad \forall c\in A\subseteq C, \\
    &   \sum_{c\in C}\vy[W,c]=1, \quad  \forall W\in\cM_k^+,\\
    & \eta(W,c)p(c,A(W,c))  +(1-\vy[W,c]) \geq\alpha, \quad \forall W\in\cM_k^+,\ c\in C,\\
    & p(c,A)\geq0, \quad \forall c\in A\subseteq C,\\
    & \vy_{W,c} \in \{0, 1\}, \quad \alpha \ge 0
\end{align*}

\begin{theorem}
\label{thm:block-marschak-equivalence}
For fixed \(m\) and \(k\), the ranking-based formulation of \infmilp~ and the Block--Marschak formulation have the same optimal value.
\end{theorem}

\begin{proof}
Consider first any feasible ranking distribution $\vx$ for \infmilp~ and define
\[
p(c,A) = \sum_{\substack{s\in\cS\\ c\succ_s A\setminus\{c\}}} \vx[s].
\]
For every nonempty $A$, these choice probabilities satisfy the normalization constraint because exactly one alternative in $A$ is highest-ranked under each strict ranking. Since they are induced by a distribution over strict rankings, their Block--Marschak quantities are nonnegative. They also reproduce exactly the same value of $P(c,W)$ for every multiset committee $W$ and alternative type $c$. So, every feasible solution of the ranking-based \infmilp~induces a feasible Block--Marschak solution with the same objective value.

Now suppose that $p(c,A)$ satisfies the normalization and Block--Marschak constraints. By the Block--Marschak representation theorem, there exists a probability distribution $\vx$ over strict rankings that induces exactly these choice probabilities. This implies that this ranking distribution produces exactly the same challenger score \(P(c,W)\) for every \(W\) and \(c\). Hence every feasible Block--Marschak solution corresponds to a feasible solution of the ranking-based \infmilp~ with the same objective value.

The two formulations therefore have identical feasible values of challenging alternatives and support, and, consequently, the same optimal value.
\end{proof}

The advantage of this reformulation is computational. The strict ranking formulation contains $m!$ variables for the rankings. In contrast, the number of stochastic choice variables is
\[ \sum_{\varnothing\neq A\subseteq C}|A| = m2^{m-1}. \]
The Block--Marschak formulation replaces the factorial number of rankings with an exponential number while preserving the optimization problem exactly. 

The same representation can also be applied to \milp. In that case, every challenger has $c\notin W$, so the relevant score simplifies to
\[
\Pr[c\succ W] = p(c,W\cup\{c\}),
\]
and no clone multiplicity factor is required. The proof of correctness follows the exact same argument as for \infmilp.

\section{Further details on \Cref{sec:axioms}}
\label{sec:appendix-axioms}

In this section, we provide additional details, including proofs and programs, for~\Cref{sec:axioms}.

\subsection{Rank-priceability}
In this section, we present the proofs of results for incomparability between stability and rank-priceability.

\PriceMILP*

\begin{proof}
For each ranking $s\in\cS$, let $N_s=\{i\in N \mid s_i=s\}$. Then $|N_s|=n\vx[s]$.

($\Rightarrow$) Suppose first that $W$ is rank-priceable, with individual payments $f_{i,w}$ and common price $\tau$. Define $ \vf[s,w]=\sum_{i\in N_s}f_{i,w}$. Summing the individual budget constraints over voters with ranking $s$ gives
$$
\sum_{w\in W}\vf[s,w] = \sum_{i\in N_s}\sum_{w\in W}f_{i,w} \leq \frac{|N_s|}{n} = \vx[s].
$$
Similarly, each winner receives
$$
\sum_{s\in\cS}\vf[s,w] = \sum_{i\in N}f_{i,w} = \tau.
$$
Finally, voters with the same ranking have the same top-$r$ alternatives. Grouping the terms in each rank constraint for each voter by ranking gives

\begin{align*}
& \sum_{i\in N \mid c\in\operatorname{top}_r(s_i)}
\left(\frac{1}{n}- \sum_{w\in W\cap\operatorname{top}_r(s_i)}f_{i,w} \right) = \\
& \sum_{s\in\cS \mid c\in\operatorname{top}_r(s)}
\left(\vx[s]- \sum_{w\in W\cap\operatorname{top}_r(s)}\vf[s,w] \right) 
\leq\tau.
\end{align*}

So all three conditions hold.

($\Leftarrow$) Suppose that $\vf$ and $\tau$ satisfy the conditions of~\Cref{lem:priceability-lp}. For each voter $i\in N_s$, distribute the aggregate payments equally among voters of that ranking,
$$ f_{i,w}=\frac{\vf[s,w]}{|N_s|}. $$
Note that $|N_s|>0$ whenever $i\in N_s$. If $\vx[s]=0$, the first condition and nonnegativity mean
$\vf[s,w]=0$ for all $w\in W$, so no payments need to be assigned to that ranking.

Each voter also respects their budget as
$$ \sum_{w\in W}f_{i,w} = \frac{\sum_{w\in W}\vf[s,w]}{|N_s|} \leq \frac{\vx[s]}{n\vx[s]} = \frac{1}{n}.
$$
The total payment to each winner is $\tau$, and grouping individual payments shows that every condition for each voter ranking is satisfied. Hence $W$ is rank-priceable.
\end{proof}

We move on to the proofs of incomparability between stability and priceability. We begin with the proof of~\Cref{thm:price-stability}. 

\PriceabilityStability*

\begin{proof}
Let $W=\{0\}$ and suppose every voter has ranking $1\succ0$.
Then, clearly $W$ is not $1$-undominated.

To see that $W$ is rank-priceable, set $\tau=1$ and let every voter spend their entire budget $1/n$ on alternative $0$. The winner is fully funded, and for alternative $1$, the rank-priceability constraint is tight at $r=1$ and trivially satisfied at $r=2$.
\end{proof}

Finally, we conclude with the proof of~\Cref{thm:stability-price}.

\StabilityPriceability*

\begin{proof}
Let $W=\{0,1\}$ and consider the rankings $s_1=(2\succ1\succ0)$ and $s_2=(1\succ2\succ0)$ with
$$\vx[s_1]=\frac{7}{12} \text{ and } \vx[s_2]=\frac{5}{12}. $$

First, $W$ is $\nicefrac{2}{3}$-undominated. Alternative $2$ is ranked above both members of $W$ only in $s_1$, and $\frac{7}{12} < \frac{2}{3}$.

Now suppose, for contradiction, that $W$ is rank-priceable with common price $\tau$. Since the total voter budget is $1$ and both committee members must each receive price $\tau$, $2\tau \leq 1 $ and therefore $ \tau\leq\frac{1}{2}$.

Now consider alternative $2$ and $r=1$. Alternative $2$ is ranked first by the voters of type $s_1$, whose total mass is $7/12$. For these voters, no member of $W$ is in their top-$1$, so none of their payments can be subtracted in the rank-priceability constraint. Hence rank-priceability requires
$ \frac{7}{12}\leq\tau$. This contradicts $\tau\leq1/2$. Therefore, $W$ is not rank-priceable.
\end{proof}

\subsection{Rank-PJR+}

In this section, we give the proofs of incomparability for Rank-PJR+ and stability as well as the proof of~\Cref{lem:pjrp-milp}.

\PJRMILP*

\begin{proof}
($\Rightarrow$) Assume $W$ satisfies Rank-PJR+. By contradiction, suppose the inequality is violated for some $c$, $r$, and $T\subsetneq W$. Then at least $(|T|+1)/k$ of the voters rank $c$ among their top $r$ alternatives while ranking no member of $W\setminus T$ among their top $r$. So, this group should have $|T|+1$ representatives in the committee, but collectively has at most the $|T|$ members of $T$ among its top-$r$ alternatives. So, $W$ violates Rank-PJR+.

($\Leftarrow$) Assume all of the inequalities above hold. By contradiction, suppose $W$ violates Rank-PJR+. Then there exist an outsider $c$, rank $r$, integer $\ell\in[k]$, and a group containing at least $\ell/k$ of the voters such that every voter ranks $c$ among their top $r$ alternatives, while the union of committee members appearing in their top-$r$ sets has size less than $\ell$. Let $T$ be this union. Then $|T|+1\leq\ell$, and every voter in the group is counted in the sum above.
Therefore,
$$
\sum_{\substack{s\in\cS \mid c\in\operatorname{top}_r(s), \\
W\cap\operatorname{top}_r(s)\subseteq T}}
\vx[s] \geq \frac{\ell}{k} \geq \frac{|T|+1}{k}.
$$
So, we obtain a contradiction.
\end{proof}

\PJRStability*
 
\begin{proof}
Let $W=\{0\}$ and consider the profile
$s_1 = (1\succ0\succ2)$ and $s_2 = (2\succ1\succ0)$ with $\vx[s_1] = \vx[s_2] = 1/2$.

Alternative $1$ is ranked above $0$ by every voter, so $W$ is not $1$-undominated.

For Rank-PJR+, since $k=1$ the only proper subset of $W$ is $T=\emptyset$, with quota $1$. For an outsider $c$ and rank $r$, we consider the fraction of voters who have $c$ within their top $r$ alternatives while the winner $0$ is not within their top $r$ alternatives. For $c=1$, this fraction is at most $1/2$, and for $c=2$ it is also at most $1/2$. Since $1/2<1$, $W$ satisfies Rank-PJR+.
\end{proof}

\StabilityPJR*

\begin{proof}
Let $W=\{0,1\}$ and suppose every voter has ranking $0\succ2\succ1$.
Alternative $2$ does not beat alternative $0$, so $W$ is $\nicefrac{2}{3}$-undominated.

However, at $r=2$, every voter ranks alternative $2$ among their top two alternatives. Since every voter ranks committee member $0$ among their top two alternatives, the electorate is represented by only one member of $W$ at this cutoff. Rank-PJR+ requires two, so $W$ does not satisfy Rank-PJR+.
\end{proof}

\subsection{Rank-EJR+}

Our framework naturally extends to additional axioms, and we give minimum incomparability results for Rank-EJR+. It is known that Rank-EJR+ implies Rank-PJR+, so our counterexample showing that Rank-EJR+ does not imply stability also applies to Rank-PJR+. We begin by recalling the definition from~\citet{Brill23:Robust}.

\begin{definition}[Rank-EJR+]
    Let $W\in\cM_k$. For a ranking $s$, let $\operatorname{top}_r(s)$ denote its top-$r$ alternatives. The committee $W$ satisfies Rank-EJR+ if, for all $r\in[m]$, all $c\notin W$, all $\ell\in[k]$, and all voter subsets $N'\subseteq N$ with
    $$  |N'|\geq \frac{\ell}{k}n, $$
    if $c\in\operatorname{top}_r(s_i)$ for all $i\in N'$ then there exists some $i\in N'$ such that
    $$
    |\operatorname{top}_r(s_i)\cap W|\geq \ell.
    $$
\end{definition}

Analogous to the two prior axioms, using our framework, we obtain the following results.  

\begin{restatable}{theorem}{RankEJRStability}[Rank-EJR$+ \not\Rightarrow$ stability]
    There exists a voter profile for $m=3$ and $k=1$ such that there is a $W \in \cM_k$ that satisfies Rank-EJR+, but is not $\alpha$-undominated for $\alpha = 1$.
\end{restatable}

\begin{proof}
Let $W=\{0\}$ and consider the rankings $s_1=(1\succ0\succ2)$ and $s_2=(2\succ1\succ0)$ with $\vx[s_1]=\vx[s_2]=1/2$.

Alternative $1$ is ranked above $0$ by every voter, so $W$ is not $1$-undominated.

For Rank-EJR+, since $k=1$, we only need to consider $\ell=1$. For either alternative $c\in\{1,2\}$ and any rank $r$, at most half of the voters rank $c$ among their top $r$ alternatives while having no member of $W$ among their top $r$ alternatives. Since
$$
\frac12 < \frac{\ell}{k}=1,
$$
no Rank-EJR+ violation exists. So, $W$ satisfies Rank-EJR+.
\end{proof}

\begin{restatable}{theorem}{StabilityRankEJR}[Stability $\not\Rightarrow$ Rank-EJR+]
    There exists a voter profile for $m=3$ and $k=2$ such that there is a $W \in \cM_k$ that is  $\alpha$-undominated for $\alpha = \nicefrac{2}{3}$, but does not satisfy Rank-EJR+.
\end{restatable}

\begin{proof}
Let $W=\{0,1\}$ and suppose every voter has the ranking $s = (1\succ2\succ0)$.

First, $W$ is $\nicefrac{2}{3}$-undominated. The only alternative not in $W$ is alternative $2$, and every voter ranks committee member $1$ above $2$.

We now show that $W$ violates Rank-EJR+. Take alternative $c=2$, $r=2$, and $\ell=2$. Every voter has
$\operatorname{top}_2(s)=\{1,2\}$,  so alternative $2$ is in every voter's top-$2$ alternatives, while
$$
|\operatorname{top}_2(s)\cap W|=1<2=\ell.
$$

Since equality with the Rank-EJR+ quota is a violation, $W$ does not satisfy Rank-EJR+.
\end{proof}

\subsection{Searching for stable committees that violate an axiom}

\subsubsection{Rank-priceability}

For a fixed profile $\vx$ and committee $W$, define the following linear program. 
\begin{align*}
 \min_{\vf,\tau,\zeta} \quad & \zeta \\
\text{s.t.} \quad & \sum_{w\in W}\vf[s,w]\leq \vx[s], \quad \forall s\in\cS, \\
& \sum_{s\in\cS}\vf[s,w]=\tau, \quad \forall w\in W, \\
& \smashoperator{\sum_{s\in\cS \mid c\in\operatorname{top}_r(s)}} \qquad \left(
    \vx[s] - \sum_{w\in W\cap\operatorname{top}_r(s)} \vf[s,w] \right) \leq \tau+\zeta, \quad \forall c\notin W, r\in[m], \\
& \vf[s,w]\geq0,\quad \tau\geq0,\quad \zeta\geq0.
\end{align*}

The variable $\zeta$ is a slack variable on the rank-priceability constraints. For any profile $\vx$ and committee $W\in\cM_k$, we get that $W$ is rank-priceable if and only if $\zeta^* = 0$. The constraints are simply imported from~\Cref{lem:priceability-lp}.

We next derive the dual of this program. Let $\vlambda[s] \geq0$ be the dual variable for the budget constraint of ranking $s$, let $\veta[w] \in \mathbb{R}$ be the dual variable for the equality constraint of committee member $w$, written as $\tau-\sum_{s\in\cS}\vf[s,w]=0$, and let $\vgamma[c,r] \geq0$ be the dual variable for the rank-priceability constraint associated with an alternative $c\notin W$ and $r\in[m]$.
Define
\begin{align*}
D_W(\vx;\vlambda,\veta,\vgamma) =  
\sum_{c\notin W}\sum_{r\in[m]} \vgamma[c,r] \sum_{\substack{s\in\cS \mid c\in\operatorname{top}_r(s)}} \vx[s]
- \sum_{s\in\cS}\vlambda[s]\vx[s].
\end{align*}

The dual program is 
\begin{align*}
\max_{\vlambda,\veta,\vgamma} \quad & D_W(\vx;\vlambda,\veta,\vgamma) \\
\text{s.t.}\quad
& \vlambda[s]-\veta[w] - \smashoperator{\sum_{c\notin W,\ r\in[m] \mid  c,w \in\operatorname{top}_r(s)} }\vgamma[c,r] \geq0,  \quad \forall s\in\cS,\ w\in W,\\
& \sum_{w\in W}\veta[w] \geq \sum_{c\notin W}\sum_{r\in[m]}\vgamma[c,r], \\
& \sum_{c\notin W}\sum_{r\in[m]}\vgamma[c,r] \leq1, \\
& 
\vlambda[s]\geq0,\qquad
\vgamma[c,r]\geq0,\qquad
\veta[w]\in\mathbb{R}.
\end{align*}

For every fixed profile $\vx$ and committee $W$, strong duality implies
that the optimal value of the primal and dual are equal. In particular, $W$ is not rank-priceable if and only if there exists a feasible solution to the dual with $D_W(\vx;\vlambda,\veta,\vgamma) > 0$.

We can use the dual program to search for profiles under which a committee is $\alpha$-undominated but not rank-priceable. Since the dual constraints are homogeneous after dropping the normalization $\sum_{c\notin W}\sum_{r\in[m]} \vgamma[c,r]\leq 1$, any feasible dual solution with strictly positive objective can be rescaled to have objective value at least $1$.

For a fixed committee $W\in\cM_k$, consider the following \textit{bilinear} optimization
problem. 
\begin{align*}
\inf_{\vx,\alpha,\vlambda,\veta,\vgamma}\quad & \alpha \\
\text{s.t.}\quad & \sum_{s\in\cS}\vx[s]=1,
\qquad \vx[s]\geq0, ~\forall s\in\cS, \\
& \smashoperator{\sum_{s\in\cS \mid c\succ_s w\ \forall w\in W}} \vx[s] \leq \alpha,~\forall c\notin W, \\
& \vlambda[s]-\veta[w] - \smashoperator{\sum_{c\notin W,\ r\in[m] \mid  c,w \in\operatorname{top}_r(s)} }\vgamma[c,r] \geq0,  \quad \forall s\in\cS,\ w\in W,\\
& \sum_{w\in W}\veta[w] \geq \sum_{c\notin W}\sum_{r\in[m]}\vgamma[c,r], \\
& D_W(\vx;\vlambda,\veta,\vgamma) \geq 1, \\
& 
\vlambda[s]\geq0,\qquad
\vgamma[c,r]\geq0,\qquad
\veta[w]\in\mathbb{R}.
\end{align*}

The first two constraints are the standard constraints from \milp~ for stability. The remaining constraints are the rank-priceability dual constraints, and adding $D_W(\vx;\vlambda,\veta,\vgamma) \ge 1$ certifies that $W$ is \textit{not} rank-priceable. Therefore, if the optimal value is strictly less than $2/(k+1)$, we obtain an instance where stability does not imply rank-priceability.

\subsubsection{Generalization to linearly verifiable axioms}
\label{sec:linear-axioms}

The preceding approach for rank-priceability extends to any axiom whose satisfaction can be verified by a linear program for a fixed profile and committee. Let $\mathcal{A}$ be such an axiom. After putting its linear constraints in standard form, suppose that for every $W\in\cM_k$,
\begin{equation*}
    W\text{ satisfies } \mathcal{A}  \Longleftrightarrow \exists \vu \text{ s.t. } A_W\vu\leq \vb_W+B_W\vx.  
\end{equation*}
Here, $\vu$ captures any additional auxiliary variables required to certify the axiom.
Equalities can be represented by additional linear constraints, so this formulation is without loss of generality.

As with rank-priceability, we introduce a common slack variable and define
the linear program
\begin{align*}
\min_{\vu,\zeta}\quad
&\zeta \\
\text{s.t.}\quad & A_W\vu\leq \vb_W+B_W\vx+\zeta\mathbf 1, \\
&\zeta\geq0.
\end{align*}
So, $W$ satisfies $\mathcal{A}$ if and only if $\zeta^* = 0$, while $\zeta^*>0$ certifies a violation of the axiom.

The dual of this program is
\begin{align*}
\max_{\vpi}\quad & D_W^{\mathcal{A}}(\vx;\vpi) := -\vpi^\top(\vb_W + B_W\vx)\\
\text{s.t.}\quad & A_W^\top\vpi=0, \\
&\mathbf{1}^\top\vpi \leq 1, \\
&\vpi\geq 0.
\end{align*}

By strong duality, $W$ does not satisfy $\mathcal{A}$ if and only if there exists $\pi$ that is feasible for the dual with $D_W^{\mathcal{A}}(\vx;\vpi)>0$. 

As before, we can drop the normalization $\mathbf 1^\top\vpi\leq1$. Define
$$ \Pi_W^{\mathcal A} = \left\{ \vpi\geq0 \mid A_W^\top\vpi=0 \right\}. $$
Any feasible dual certificate with strictly positive objective can be rescaled so that its objective is at least one. 

We obtain a general optimization program for searching for $\alpha$-undominated committees that violate $\mathcal{A}$. For a fixed committee $W$, consider
\begin{align*}
\inf_{\vx,\alpha,\vpi}\quad
&\alpha
\label{eq:stable-not-axiom}\\
\text{s.t.}\quad
& \sum_{s\in \cS} \vx[s]= 1,\\
& \smashoperator{\sum_{\csuccW}} \vx[s] \leq\alpha, \forall~c\notin W, \\
&\vpi\in\Pi_W^{\mathcal A}, \\
& D_W^{\mathcal A}(\vx;\vpi) \geq 1, \\
& \vx[s] \ge 0, \quad \alpha\geq0.
\end{align*}
The objective minimizes the largest fraction of voters supporting an alternative against $W$, while the final two constraints certify that $W$ violates $\mathcal{A}$.
The resulting program is generally bilinear, since $D_W^{\mathcal A}(\vx;\vpi)$ contains products between the profile variables and the dual variables. Rank-priceability is an instance of this general construction, with its payment variables providing the primal certificate
and $(\vlambda,\veta,\vgamma)$ providing the dual certificate. 

\subsubsection{Rank-PJR+}

The corresponding program for Rank-PJR+ is actually simpler than the above construction. By \Cref{lem:pjrp-milp}, a violation of Rank-PJR+ is already characterized by a linear inequality depending only on the profile $\vx$. So, unlike rank-priceability, we do not need to introduce an auxiliary feasibility program or dualize its constraints.

For a fixed committee $W\in\cM_k$ and fixed $c\notin W$, $r\in[m]$, and
$T\subsetneq W$, consider
\begin{align*}
\min_{\vx,\alpha} \quad & \alpha \\
\text{s.t.}\quad & \sum_{s\in\cS}\vx[s]=1, \\
& \smashoperator{\sum_{s\in\cS \mid c'\succ_s w\ \forall w\in W}} \vx[s]\leq\alpha, ~\forall c'\notin W,\\
& \smashoperator{\sum_{\substack{s\in\cS \mid
c\in\operatorname{top}_r(s),\\
W\cap\operatorname{top}_r(s)\subseteq T}}}
\vx[s] \geq\frac{|T|+1}{k}. \\
& 
\vx[s] \ge 0.
\end{align*}

The first two constraints minimize the largest fraction of voters supporting an alternative to $W$, while the final constraint certifies a Rank-PJR+ violation. We solve the program over all choices of $c$, $r$, and $T$. If the minimum objective value is strictly less than $2/(k+1)$, then there exists a $2/(k+1)$-undominated committee that does not satisfy Rank-PJR+.

An analogous program can be obtained for Rank-EJR+ by replacing the final constraint with its corresponding linear violation condition.

\subsection{Searching for axiom-satisfying committees that violate stability}

Finally, we consider the reverse implications of incomparability for a general axiom $\mathcal{A}$. Assume again that the satisfaction of $\mathcal{A}$ can be determined by a linear system for a fixed profile $\vx$ and committee $W$,
$$
A_W\vu\leq \vb_W+B_W\vx,
$$
for some auxiliary variables $\vu$. Unlike the previous direction, we do not need to dualize these constraints, since here we require $W$ to satisfy the axiom.

For a fixed committee $W\in\cM_k$ and alternative $c\notin W$, consider
\begin{align*}
\max_{\vx,\vu} \quad
& \sum_{\substack{s\in\cS \mid c\succ_s w\ \forall w\in W}}\vx[s]\\
\text{s.t.}\quad
& \sum_{s\in\cS}\vx[s]=1,\\
& A_W\vu\leq \vb_W+B_W\vx, \\
& \vx[s]\geq0.
\end{align*}

The constraints require $W$ to satisfy $\mathcal{A}$, while the objective maximizes the fraction of voters for which $c$ is preferred to every member of $W$.  If the optimal value is
at least $\nicefrac{2}{(k+1)}$, then there exists a committee satisfying $\mathcal{A}$ that is not
$\nicefrac{2}{(k+1)}$-undominated. 

For axioms whose satisfaction is characterized by strict linear inequalities, such as Rank-PJR+, we introduce a slack variable $\epsilon\geq0$ and write
$$ A_W\vu+\epsilon\mathbf 1\leq \vb_W+B_W\vx. $$
The axiom is satisfied if and only if there exists a feasible solution with $\epsilon>0$. For a fixed threshold $\alpha$, the resulting program maximizes $\epsilon$ subject to the axiom constraints and the requirement that $W$ is not $\alpha$-undominated. A strictly positive optimum certifies that $W$ satisfies the axiom while violating $\alpha$-undomination.

\section{Omitted tables}
\label{sec:omitted-tables}

\setcounter{table}{0}
\renewcommand{\thetable}{A\arabic{table}}

\begin{table*}[h!]
    \centering
    \caption{Max $\alpha$ values from $\texttt{infMILP}$ with an 81,400 s timeout. The final row gives the reference threshold $\nicefrac{2}{(k+1)}$.}
    \label{tab:infmilp-results-full}
    \begin{tabularx}{\linewidth}{c|Y|Y|Y|Y|Y}
        \toprule
        \textbf{m} \textbackslash ~\textbf{k} & \textbf{1} & \textbf{2} & \textbf{3} & \textbf{4} & \textbf{5}\\
         \midrule
        \textbf{2}  & 1.00 & -- & --  &  -- & -- \\
       \textbf{3} & 1.00 & 0.667 & --  &  -- & -- \\
       \textbf{4}   & 1.00 & 0.667 & 0.500 & -- & -- \\
       \textbf{5}  & 1.00 & 0.667 & 0.500 &  0.400 & -- \\
       \textbf{6}   & 1.00 & 0.667 &  0.500  & [0.400, 0.500] & [0.333, 0.429] \\
       \textbf{7}   & 1.00 & 0.667 & 0.500 & [0.400, 0.500] & [0.333, 0.500] \\
       \textbf{8}   & 1.00 & [0.667, 0.750] & [0.500, 0.667] & [0.400, 0.667] & [0.333, 0.667] \\
        \textbf{9} & 1.00 & [0.667, 0.800]  & [0.500, 0.750] & [0.400, 0.750] & [0.333, 0.925]\\
        \midrule
       $\mathbf{\nicefrac{\mathbf{2}}{(k+1)}}$ & 1.00 & 0.667 & 0.500 & 0.400 & 0.333 \\ 
       \bottomrule
    \end{tabularx}
    
\end{table*}

\end{document}